\documentclass[fleqn,usenatbib]{mnras}
\usepackage{newtxtext,newtxmath}
\usepackage[T1]{fontenc}
\usepackage{ae,aecompl}

\usepackage{graphicx}	
\usepackage{amsmath}	
\usepackage{subcaption}
\usepackage{bm}
\usepackage[utf8]{inputenc}

\title[Feedback from an O star formed in a filament]{Ionising feedback from an O star formed in a filament}
\author[A. P. Whitworth \& F. D. Priestley]{A. P. Whitworth$^{1}$\thanks{E-mail: ant@astro.cf.ac.uk}
and F. D. Priestley$^{1}$\\
$^{1}$School of Physics and Astronomy, Cardiff University, Cardiff CF24 3AA, UK}
\date{Accepted XXX. Received YYY; in original form ZZZ}
\pubyear{2019}
\begin{document}
\label{firstpage}
\pagerange{\pageref{firstpage}--\pageref{lastpage}}
\maketitle

\begin{abstract}
We explore a simple semi-analytic model for what happens when an O star (or cluster of O stars) forms in an isolated filamentary cloud. The model is characterised by three configuration parameters: the radius of the filament, $R_{_{\rm FIL}}$, the mean density of H$_2$ in the filament, $n_{_{\rm FIL}}$, and the rate at which the O star emits ionising photons, $\dot{\cal N}_{_{\rm LyC}}$. We show that for a wide range of these configuration parameters, ionising radiation from the O star rapidly erodes the filament, and the ionised gas from the filament disperses into the surroundings. Under these circumstances the distance, $L$, from the O star to the ionisation front (IF) is given approximately by $L(t)$$\,\sim$$\,5.2\,{\rm pc}$$\,[R_{_{\rm FIL}}/0.2\,{\rm pc}]^{-1/6}$$\,[n_{_{\rm FIL}}/10^4\,{\rm cm^{-3}}]^{-1/3}$$\,[\dot{\cal N}_{_{\rm LyC}}/10^{49}\,{\rm s}^{-1}]^{1/6}$$\,[t/{\rm Myr}]^{2/3}$, and we derive similar simple power-law expressions for other quantities, for example the rate at which ionised gas boils off the filament, $\dot{M}_{_{\rm IF}}(t)$, and the mass, $M_{_{\rm SCL}}(t)$, of the shock-compressed layer (SCL) that is swept up behind the IF. We show that a very small fraction of the ionising radiation is expended locally, and a rather small amount of molecular gas is ionised and dispersed. We discuss some features of more realistic models, and the extent to which they might modify or invalidate the predictions of this idealised model. In particular we show that, for very large $R_{_{\rm FIL}}$ and/or large $n_{_{\rm FIL}}$ and/or low $\dot{\cal N}_{_{\rm LyC}}$, continuing accretion onto the filament might trap the ionising radiation from the O star, slowing erosion of the filament even further. 
\end{abstract}


\section{Context}

Feedback from high-mass stars is a critical process in astrophysics, firstly because it determines the external appearance of the resulting nebulae (H{\sc ii} regions, stellar-wind bubbles, supernova remnants); secondly because it is presumed to play an important role in the self-regulation of star formation; and thirdly because this self-regulation is a key element moderating the formation, structure and evolution of galaxies, and their interaction with the intergalactic medium \citep[e.g.][]{ChevanceMetal2020}.

However, the detailed mechanics of feedback is very complicated, due to the non-linear nature of the processes involved, and the complexity of the initial conditions \citep{DaleJE2015}. Robust theorems quantifying the effectiveness of feedback in triggering further star formation, the rate at which feedback disperses star-forming molecular clouds, and the fraction of ionising radiation that escapes the immediate vicinity of a molecular cloud, are hard to formulate, and their are predictions sometimes contradictory.

Under certain circumstances ionising radiation may act to promote star formation, for example when expansion of an H{\sc ii} region sweeps up gas into a dense shell or layer which then fragments to produce a new generation of stars \citep[e.g.][this process is often termed {\it collect and collapse}]{ElmegreenLada1977, ElmegreeBGDM1978, Whitetal1994a, Whitetal1994b, DaleJetal2007}. Under other circumstances, an expanding H{\sc ii} region may form and compresses a more local density enhancement, for example a pillar \citep[e.g.][]{GritschnederMetal2010, TremblinPetal2012a, TremblinPetal2012b}, or a globule \citep[e.g.][this process if often termed {\it radiatively driven implosion}]{BertoldiF1989, LeflochLazareff1994, BisbasTetal2011}. However, when the pre-existing internal structure of a molecular cloud is taken into account, it appears that these processes may simply accelerate star formation, and that in the long term the net amount of star formation may actually be reduced \citep[e.g.][]{DaleBonnell2011, WalchSetal2012, WalchSetal2013, HaidSetal2019} --- in the sense that, if they had not occurred, a larger mass of stars would have formed, just on a longer timescale. 

A key issue is the distribution of gas and dust surrounding a newly-formed massive star or star cluster, since this influences the extent to which the radiation and high-energy (or high-momentum) gas excited by the stars is trapped locally. At one extreme, if one considers an H{\sc ii} region at the centre of a collapsing cloud, the H{\sc ii} region may be trapped if the cloud is sufficiently massive \citep[e.g.][]{DaleJEetal2012, TremblinPetal2014, GeenSetal2015}. Conversely, if an H{\sc ii} region forms close to the boundary of a molecular cloud, it will break out rather easily, dispersing part of the cloud and releasing ionising radiation into the surroundings \citep[e.g.][]{WhitworthA1979, MatznerCD2002}, and even if the ionising stars are located near the centre of the cloud, the ionising radiation and hot gas may escape quite freely through channels between the denser star-forming gas \citep[e.g.][]{DaleJEetal2012}. In the present paper we explore the situation where stars form in a filament. We show that if the filament is isolated, or if accretion onto the filament is petering out, then very little molecular gas is ionised and dispersed, much more molecular gas is swept up into a dense layer, and a very large fraction of the ionising radiation escapes.

\section{Introduction}\label{SEC:Intro}

It appears that -- at least in the local Universe -- many stars form in filaments. In some star forming regions, there are several separate filaments, each seemingly spawning stars independently, and producing a distributed population of stars, as in Ophiuchus or Taurus \citep[e.g.][]{AndrePhetal2010, AndrePetal2014, Arzoumanetal2019, HowardADetal2019, Ladjelatetal2020}. In other star forming regions, there is a single massive filament, as in G316.75-00.00 \citep{WatkinsEetal2019} and RCW36 \citep{MinierVetal2013}, or a system of filaments radiating from a central massive core (a {\it hub-and-spoke} system), as in SDC335.579-0.272 \citep{PerettoNetal2013}, SCD13.174 \citep{WilliamsG2018}, the Orion Molecular Cloud \citep{HacarAetal2018}, the Rosette Molecular Cloud \citep{Schneideretal2012}, and MonR2 \citep{TrevinoMetal2019}. These are the filaments that seem often to support the formation of star clusters and massive stars. Therefore it is important to understand how feedback from a massive star formed in such a filament subsequently affects the structure and evolution of the filament.

\begin{figure*}
\vspace*{-4.0cm}\hspace*{-2.50cm}\includegraphics[angle=-90.,width=2.75\columnwidth]{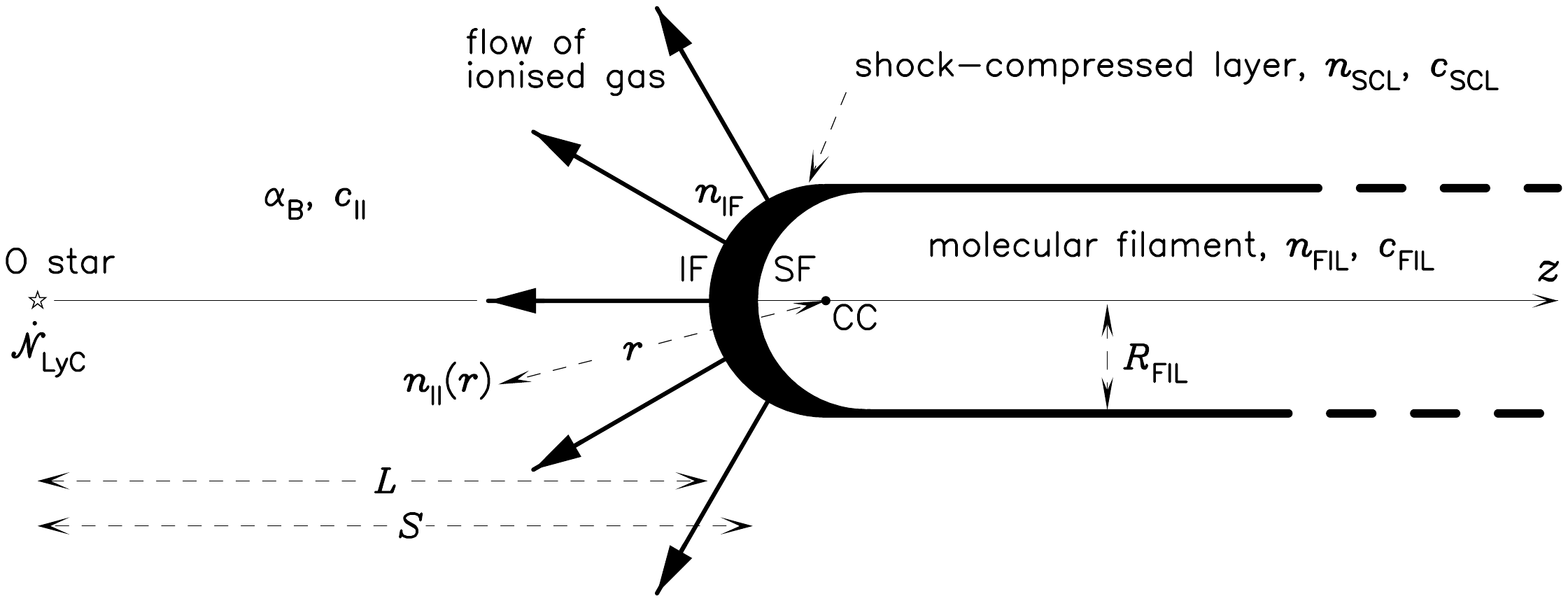}
\vspace*{-4.5cm}
\caption{Cartoon illustrating the erosion of a filament by an O star. The bold outline on the right represents the boundary of a cylindrically symmetric filament, characterised by its radius, $R_{_{\rm FIL}}$, its uniform molecular-hydrogen density, $n_{_{\rm FIL}}$ and its uniform velocity dispersion, $\sigma_{_{\rm FIL}}$. On the left is an O star which emits ionising photons at rate $\dot{\cal N}_{_{\rm LyC}}$, and drives an ionisation front (IF) into the end of the filament. The O star is on the axis of symmetry of the filament (the $z$ axis), because it has formed out of the filament. The IF has radius of curvature $R_{_{\rm FIL}}$, and centre of curvature (CC) at distance $L+R_{_{\rm FIL}}$ from the O star. Ionised gas flows off the IF normally, as indicated by the solid radial arrows, and at the adiabatic sound speed in the ionised gas, $c_{_{\rm II}}$. The density of protons at the IF, on the ionised side, is $n_{\rm II}(R_{_{\rm IF}})\!=\!n_{_{\rm IF}}$, and it decreases with distance, $r$, from the CC as $n_{\rm II}(r)\sim r^{-2}$ (see Eq.\ref{EQN:npr.1}). The gas-kinetic temperature in the ionised gas is uniform and constant, and hence the recombination coefficient, $\alpha_{_{\rm B}}$, and the adiabatic sound speed, $c_{_{\rm II}}$, are also uniform and constant. Behind the IF is a shock front (SF), at distance $S$ from the O star. Between the IF and the SF is a shock-compressed layer (SCL) of neutral gas. The SCL has thickness $W=S-L$, and is represented by the meniscus-shaped solid black region.}
\label{FIG:Cartoon_FiF}
\end{figure*}

In \S\ref{SEC:Model}, we develop a simple geometric model for the situation where the filament is eroded by the ionising radiation from the O star and disperses freely into the surroundings (see Fig. \ref{FIG:Cartoon_FiF}); this model forms the basis for the subsequenty sections (\ref{SEC:HIIRionbal} through \ref{SEC:Approximations}). In \S\ref{SEC:HIIRionbal} we analyse the structure of the resulting H{\sc ii} region; we focus on the attenuation of ionising radiation on the line of sight between the massive star and the ionisation front (IF) where the ionising radiation is boiling material off the exposed end of the filament. In \S\ref{SEC:SCLstructure} we analyse the structure of the thin shock-compressed layer (SCL) that forms immediately behind the IF, i.e. between the IF and the shock front (SF) driven into the undisturbed neutral gas of the filament, ahead of the IF. In \S\ref{SEC:IFlocation} we derive an expression for the location of the IF, as a function of time. In \S\ref{HIIRdynamics} we derive expressions for the rate at which mass boils off the filament. In \S\ref{SEC:SCLgrowth} we derive expressions for the global properties of the shock-compressed layer between the IF and the SF. In \S\ref{SEC:Approximations} we review the approximations made in our analysis, and evaluate the assumptions underlying the model. In \S\ref{SEC:trappedHIIregion} we evaluate the alternative situation where continuing accretion onto the filament traps the ionised gas, so it is unable to disperse, and erosion of the filament stalls. In \S\ref{SEC:Conclusions} we discuss the results and summarise our conclusions. We neglect the role of magnetic fields.

\section{Model}\label{SEC:Model}

We model the filament as a semi-infinite homogeneous cylinder whose axis of symmetry coincides with the $z$ axis (see Fig. \ref{FIG:Cartoon_FiF}). The filament has radius $R_{_{\rm FIL}}$, and we define a dimensionless variable
\begin{eqnarray}
R_{_{0.2}}&=&\frac{R_{_{\rm FIL}}}{0.2\,{\rm pc}}\,.
\end{eqnarray}
The gas has solar composition ($X=0.70,\, Y=0.28,\, Z=0.02$), and we assume that in the filament all the hydrogen is molecular. The mean mass per hydrogen molecule, when other elements (in particular helium) are taken into account, is therefore ${\bar m}_{_{\rm H_2}}\!=\!2m_{_{\rm H}}/X\!=\!4.75\times 10^{-24}\,{\rm g}$ (where $m_{_{\rm H}}$ is the mass of an hydrogen atom). Similarly, in the H{\sc ii} region the mean mass per proton is $\bar{m}_{_{\rm H}}\!=\!m_{_{\rm H}}/X\!=\!2.37\times 10^{-24}\,{\rm g}$. In the undisturbed filament the density of molecular hydrogen is  $n_{_{\rm FIL}}$, and we define a dimensionless variable
\begin{eqnarray}
n_{_4}&=&\frac{n_{_{\rm FIL}}}{10^4\,{\rm H_{_2}\,cm^{-3}}}\,.
\end{eqnarray}
The line density of the undisturbed filament,
\begin{eqnarray}
\mu_{_{\rm FIL}}&=&\pi R_{_{\rm FIL}}^2n_{_{\rm FIL}}{\bar m}_{_{\rm H_2}}\\
&\rightarrow&88\,{\rm M\,pc^{-1}}\;R_{_{0.2}}^2\,n_{_4}\,,
\end{eqnarray}
is assumed to be critical (in the sense that it is just supported against self-gravity), so the effective sound speed, representing thermal and non-thermal (turbulent) motions, is
\begin{eqnarray}\label{EQN:cFIL.1}
c_{_{\rm FIL}}&=&\left[\pi Gn_{_{\rm FIL}}{\bar m}_{_{\rm H_2}}/2\right]^{1/2}\,R_{_{\rm FIL}}\\\label{EQN:cFIL.2}
&\rightarrow&0.435\,{\rm km\,s^{-1}}\;R_{_{0.2}}\,n_{_4}^{1/2}.
\end{eqnarray}

We assume that an O star, or compact group of OB stars (hereafter `the O star'), has formed from the filament, and is located on the $z$ axis at $[x,y,z]=[0,0,0]$. At time $t\!=\!0$ the O star starts to emit ionising photons at a constant rate $\dot{\cal N}_{_{\rm LyC}}$, and we define a dimensionless variable
\begin{eqnarray}
\dot{\cal N}_{_{49}}&=&\frac{\dot{\cal N}_{_{\rm LyC}}}{10^{49}\,{\rm s^{-1}}}.
\end{eqnarray}
As a result, a section of the filament becomes ionised and the ionised gas streams away from the filament and disperses  (see Fig. \ref{FIG:Cartoon_FiF}). Consequently the ionising radiation becomes less strongly attenuated and the IF advances into the filament.

The IF driven into the exposed end of the filament is preceded by a shock front (SF), which sweeps up the the neutral gas of the filament into a shock-compressed layer (SCL). The passage of the SF will not change the gas temperature much, but it will amplify the turbulent motions, so that the effective sound speed in the shock-compressed layer is increased somewhat, and we assume that $c_{_{\rm SCL}}\!\approx 2^{1/2}c_{_{\rm FIL}}$ (see Eq. \ref{EQN:cSCL.1}); in other words the amount of turbulent energy is doubled.\footnote{In the situation considered here, the amplification of turbulence by the shock is attributable to the Non-Linear Thin-Shell instability \citep{VishniacE1994} and Kelvin-Helmholtz Instability \citep[e.g.][]{HeitschFetal2005}. However, there is no simple analytic expression for the amplification factor, $f$, and we adopt $f\!\rightarrow\!2^{1/2}$ as indicative of the sort of values to be expected. It is straightforward to include $f$ as a fourth configuration parameter. However, given the small expected range of $f$ (as compared with $R_{_{\rm FIL}}$, $n_{_{\rm FIL}}$ and $\dot{\cal N}_{_{\rm LyC}}$), and the fact that $f$ has a very weak power-law dependence in the final expressions (Eqs. 87 to 98), we have omitted it, on the grounds that the other dependences are much more significant.}

\begin{table}
\begin{center}
\caption{Definitions of acronyms and mathematical symbols.}
\begin{tabular}{lll}\hline
\multicolumn{3}{l}{\sc Acronyms} \\
H{\sc ii} region & H{\sc ii}R & \\
Ionisation front & IF & \\
Centre of curvature of IF & CC & \\
Shock front & SF & \\
Shock-compressed layer & SCL & \\\hline
\multicolumn{3}{l}{\sc Configuration Parameters and Fiducial Values} \\
Radius of filament & $R_{_{\rm FIL}}$ & $=R_{_{0.2}}\times\left[0.2\,{\rm pc}\right]$ \\
Density of H$_2$ in filament & $n_{_{\rm FIL}}$ & $=n_{_{4}}\times\left[10^4\,\rm{cm}^{-3}\right]$ \\
Output of ionising photons & $\dot{\cal N}_{_{\rm LyC}}$ & $=\dot{\cal N}_{_{49}}\,\times\left[10^{49}\,\rm{s}^{-1}\right]$ \\\hline
\multicolumn{3}{l}{\sc Fixed Parameters and Values} \\
Temperature in ionised gas & $T_{_{\rm II}}$ & $10^4\,{\rm K}$ \\
Sound speed in ionised gas & $c_{_{\rm II}}$ & $15\,{\rm km\,s^{-1}}$ \\
Case B recombination coefficent & $\alpha_{_{\rm B}}$ & $2.7\!\times\!10^{-13}\rm{cm^3s^{-1}}$ \\
Effective sound speed in filament & $c_{_{\rm FIL}}$ & $0.435\,{\rm km\,s^{-1}}R_{_{0.2}}n_{_4}^{1/2}$ \\
Effective sound speed in SCL & $c_{_{\rm SCL}}$ & $0.615\,{\rm km\,s^{-1}}R_{_{0.2}}n_{_4}^{1/2}$ \\\hline
\multicolumn{3}{l}{\sc Dependent Variables} \\
Distance from O star to IF & $L(t)$  & \\
Distance from O star to SF & $S(t)$ & \\
Thickness of SCL & $W(t)$ & $=S(t)\!-\!L(t)$ \\
Density of protons at IF & $n_{_{\rm IF}}(t)$ & \\
Density of protons in H{\sc ii}R & $n_{\rm II}(r,t)$ & \\
Density of H$_2$ in SCL & $n_{_{\rm SCL}}(t)$ & \\
Bulk velocity in SCL & $v_{_{\rm SCL}}(t)$ & \\
Mass of SCL & $M_{_{\rm SCL}}\!(t)$ & \\\hline
\multicolumn{3}{l}{\sc Independent Variables} \\
Time since O star switch-on \;\;& $t$ & \\
Radial distance from CC & $r$ & \\
Distance from O star on & & \\
\hspace{1.8cm} symmetry axis & $z$ & \\\hline
\multicolumn{3}{l}{\sc Dimensionless Variables} \\
Dimensionless $L$ & $\lambda$ & $=L/R_{_{\rm FIL}}$ \\
Dimensionless $W$ & $\omega$ & $=W/R_{_{\rm FIL}}$ \\
Dimensionless $t$ & $\tau$ & $=c_{_{\rm FIL}}t/R_{_{\rm FIL}}$ \\\hline
\end{tabular}
\end{center}
\label{TAB:Params}
\end{table}

We assume that the ionised gas has uniform and constant gas-kinetic temperature, $T_{_{\rm II}}=10^4\,{\rm K}$, and hence uniform and constant adiabatic sound speed, $c_{_{\rm II}}=15\,{\rm km\,s^{-1}}$, and uniform and constant Case B recombination coefficient, $\alpha_{_{\rm B}}=2.7\times 10^{-13}\,\rm{cm^3\,s^{-1}}$. Case B recombination invokes the On-the-Spot Approximation: recombinations straight into the ground state of hydrogen are ignored, on the assumption that such recombinations result in the emission of photons just above the Lyman continuum limit, where the cross-section presented by hydrogen atoms is large, and so these photons are unlikely to travel far before producing compensatory ionisations. We also neglect ionisation of helium. The assumption of uniform and constant temperature, and the neglect of helium ionisation, are reasonable because we shall be mainly concerned with the ionised gas near the IF. Here the hydrogen-ionising radiation is rather hard and likely to maintain a relatively high gas-kinetic temperature, $T_{_{\rm II}}\!\sim\!10^4\,{\rm K}$. Moreover, unless the O star is exceptionally hot, no helium-ionising photons will reach that far, i.e. the helium ionisation fronts will be closer to the O star.

At the heart of our model are the following two {\it Ans{\"a}tze}.\\
\noindent {\it Ansatz 1.} The ionised gas flowing off the exposed end of the filament diverges, and so the recombination rate decreases with distance from the IF. This divergence of the flow is largely determined by the curvature of the IF, and we assume that this curvature is of order $R_{_{\rm FIL}}^{-1}$. This enables us to compute the position of the IF by integrating the rate of recombination along the $z$ axis from the O star to the centre of the IF at $[x,y,z]=[0,0,L(t)]$ (see Fig. \ref{FIG:Cartoon_FiF} and Eq. \ref{EQN:IonBal.2}).

\noindent {\it Ansatz 2.} Ionised gas flows off each point on the IF at the same rate, and estimate this rate using equations derived for the central point on the IF at $[x,y,z]=[0,0,L(t)]$. Evidently this overestimates the flow-rate off all other points on the IF -- by a factor that increases with distance from the $z$ axis, in particular because ionising radiation arrives at these points at an increasingly oblique angle to the IF. We compensate for this overestimate by assuming that the area of the IF is only $\pi R_{_{\rm FIL}}^2$ (i.e. the cross-section of the filament, as seen from the O star). The IF is therefore {\it tetartospherical}.\footnote{We use the prefix `tetarto' ($\tau\!\epsilon\!\tau\!\alpha\!\rho\!\tau\!{o}$) in `tetartospherical' to designate a quarter of a sphere. We use a Greek prefix to be consistent with the standard nomenclature, `hemispherical'.} 

The centre of curvature of the IF (hereafter the CC) is therefore the point $[x,y,z]=[0,0,L(t)\!+\!R_{_{\rm FIL}}]$, and the IF is located at
\begin{eqnarray}\label{EQN:IF.1}
z\!&\!\approx\!&\!L(t)+\left[x^2+y^2\right]^{1/2}\!,\hspace{0.9cm}x^2+y^2\,<\,3R_{_{\rm FIL}}^2/4\,,\hspace{0.7cm}
\end{eqnarray}
where the positive square-root should be taken.

The IF is preceded by a shock front (SF), which sweeps up the neutral gas in the filament and compresses it, so that it has the appropriate density for a D-critical IF. We assume that the SF sits close behind the IF, with position
\begin{eqnarray}
z\!&\!\approx\!&\!S(t)+\left[x^2+y^2\right]^{1/2}\!,\hspace{0.9cm}x^2+y^2\,<\,3R_{_{\rm FIL}}^2/4\,,\hspace{0.7cm}
\end{eqnarray}
where $S(t)$ is the distance along the $z$ axis from the O star to the SF, and again the positive square-root should be taken. We also define
\begin{eqnarray}\label{EQN:W.1}
W(t)&=&S(t)\,-\,L(t)\,,
\end{eqnarray}
which is the thickness of the SCL between the IF and the SF. In general, $W(t)\ll L(t)$ and hence $S(t)\sim L(t)$, as shown in  \S\ref{SEC:SCLgrowth}.

We assume that the O star has formed near the end of a filament; in \S\ref{SUBSEC:Ostarinthemiddle} we discuss the extent to which the model needs to be adjusted if the O star forms somewhere near the middle of the filament.

\section{Ionisation balance}\label{SEC:HIIRionbal}

For the sake of simplicity, we exploit the fact that there is a rather short period when (a) the gas has not had time to move far, (b) most of the ionising photons are expended ionising gas for the first time, and (c) the IF is R-type \citep{KahnFD1954}. This period is of order a few (say ten) recombination times,
\begin{eqnarray} 
0\;\;<\;\,t\;\;\la\;\;\frac{5\;\;}{\alpha_{_{\rm B}}\!(T)\,n_{_{\rm FIL}}}\!&\!\rightarrow\!&\!0.00006\,\rm{Myr}\;\;n_{_4}^{-1}\,.\hspace{0.5cm}
\end{eqnarray}
After this the IF switches to being D-critical \citep{KahnFD1954}, and most of the ionising radiation is expended maintaining ionisation against recombination in the region between the O star and the IF. Only a small fraction of the ionising radiation gets to ionise new material at the advancing IF (we check this retrospectively in \S\ref{SUBSEC:checkIFfraction}). The newly ionised gas is significantly over-pressured, and therefore expands away from the filament.

Ionised gas flows off the D-critical IF at the adiabatic sound speed, $c_{_{\rm II}}$, and we assume that the volume-density of protons in the ionised gas, $n_{\rm p}$ decreases approximately as $(r/R_{_{\rm FIL}})^{-2}$, where $r$ is distance from the CC. The outward flow of ionised gas is accelerated somewhat by the inward pressure gradient, and therefore on this count the density of protons will decrease even faster than $(r/R_{_{\rm FIL}})^{-2}$. On the other hand, the curvature of the IF is probably a little larger than $R_{_{\rm FIL}}$, and so on this count the density of protons should decreases somewhat slower than $(r/R_{_{\rm FIL}})^{-2}$, at least near the IF. A proper evaluation of the competition between these two effects lies outside the scope of our model. Here we simply posit that the key feature of the flow of ionised gas off the end of a filament is that the divergence of the flow is characterised by a length-scale of order the radius of the filament, $R_{_{\rm FIL}}$ ({\it Ansatz 1}). In the remainder of this section and the following four sections (\S\ref{SEC:SCLstructure} through \S\ref{SEC:SCLgrowth}) we focus on estimating conditions on the axis of symmetry (the $z$ axis). 

Since we are assuming that the O star is not sufficiently hot to ionise helium all the way to the IF, the volume-density of protons is the same as the volume-density of electrons and given by
\begin{eqnarray}\label{EQN:npr.1}
n_{_{\rm II}}(r,t)&\approx&n_{_{\rm IF}}(t)\,\left[\!\frac{r}{R_{_{\rm FIL}}}\!\right]^{-2}\,;\hspace{0.5cm}
\end{eqnarray}
the recombination rate per unit volume is therefore
\begin{eqnarray}\label{EQN:RecombRate}
{\cal R}(r,t)&=&\alpha_{_{\rm B}}\,n_{_{\rm II}}^2(r,t)\;\;\approx\;\;\alpha_{_{\rm B}}\,n_{_{\rm IF}}^2(t)\,\left[\!\frac{r}{R_{_{\rm FIL}}}\!\right]^{-\,4}.\hspace{0.6cm}
\end{eqnarray}
Here $n_{_{\rm IF}}(t)$ is the volume-density of protons immediately outside the tetartospherical IF. It is acceptable to split the dependence on $t$ and $r$ in the way we have done, provided that the IF advances on a timescale much longer than the timescale it takes for newly ionised gas to get far from the IF, i.e.
\begin{eqnarray}\label{INEQ:subsonicIFSF}
\frac{L(t)}{dL/dt}&\gg&\frac{R_{_{\rm FIL}}}{c_{_{\rm II}}}\,;
\end{eqnarray}
we check this retrospectively in \S\ref{SUBSEC:checkslowIFadvance}.

On the $z$ axis, we have $r\!=\!L(t)\!+\!R_{_{\rm FIL}}\!\!-\!z$. Therefore, if we (a) neglect the small fraction of ionising photons that reaches the ionisation front and ionises new material, and (b) equate the supply of ionising photons to the rate of recombination integrated along the $z$ axis from the O star to the IF, ionisation balance requires
\begin{eqnarray}\label{EQN:IonBal.1}
\dot{\cal N}_{_{\rm LyC}}\!\!&\!\!\approx\!\!&\!\!\int\limits_{z=0}^{z=L(t)}\frac{\alpha_{_{\rm B}}\,n_{_{\rm IF}}^2\!(t)\,R_{_{\rm FIL}}^4\,4\,\pi\,z^2\,dz}{\left[L(t)+R_{_{\rm FIL}}-z\right]^4}\\\label{EQN:IonBal.2}
\!\!&\!\!\approx\!\!&\!\!\frac{4\pi\alpha_{_{\rm B}}n_{_{\rm IF}}^2(t)R_{_{\rm FIL}}^3}{3}\!\left\{\frac{L^2(t)}{R_{_{\rm FIL}}^2}\!-\!\frac{L(t)}{R_{_{\rm FIL}}}\!+\!1\!-\!\frac{R_{_{\rm FIL}}}{[R_{_{\rm FIL}}\!+\!L(t)]}\right\}\!\!.\hspace{0.5cm}
\end{eqnarray}

\section{Structure of the shock-compressed layer}\label{SEC:SCLstructure}

The density of molecular hydrogen in the SCL is
\begin{eqnarray}\label{EQN:shock.1}
n_{_{\rm SCL}}\!(t)&\approx&n_{_{\rm FIL}}\left[\frac{dS\!/\!dt}{c_{_{\rm SCL}}}\right]^{\,2}\,,
\end{eqnarray} 
and conservation of mass requires 
\begin{eqnarray}\label{EQN:shock.2}
n_{_{\rm FIL}}\,dS\!/\!dt&\approx&n_{_{\rm SCL}}\!(t)\,\left[dS\!/\!dt\,-\,\upsilon_{_{\rm SCL}}\!(t)\right]\,,
\end{eqnarray}
where $\upsilon_{_{\rm SCL}}\!(t)$ is the velocity of the shock-compressed gas, parallel to the $z$ axis. Eliminating $n_{_{\rm SCL}}\!(t)$ between Eqs. \ref{EQN:shock.1} and \ref{EQN:shock.2}, we obtain
\begin{eqnarray}\label{EQN:vSCL.1}
\upsilon_{_{\rm SCL}}\!(t)&\approx&dS\!/\!dt\,-\,\frac{c_{_{\rm SCL}}^2}{dS\!/\!dt}\,.
\end{eqnarray}

As shown in the Appendix, material flows into the D-critical IF at speed
\begin{eqnarray}\label{EQN:dLdt.1}
\frac{dL}{dt}\,-\,\upsilon_{_{\rm SCL}}\!(t)&\approx&\frac{c_{_{\rm SCL}}^2}{2c_{_{\rm II}}}\,,
\end{eqnarray}
and out at speed $c_{_{\rm II}}$, so conservation of mass across the IF gives
\begin{eqnarray}\label{EQN:DcIF.1}
n_{_{\rm SCL}}\!(t)&\approx&n_{_{\rm IF}}(t)\left[c_{_{\rm II}}/c_{_{\rm SCL}}\right]^2\,.
\end{eqnarray}
Eliminating $n_{_{\rm SCL}}\!(t)$ between Eqs. \ref{EQN:shock.1} and \ref{EQN:DcIF.1}, we obtain
\begin{eqnarray}\label{EQN:dSdt.1}
dS\!/\!dt&\approx&c_{_{\rm II}}\left[n_{_{\rm IF}}(t)/n_{_{\rm FIL}}\right]^{1/2}\,.
\end{eqnarray}

If we now define the dimensionless parameter,
\begin{eqnarray}\label{EQN:chi.1}
\chi(t)\!&\!=\!&\!\left[\frac{n_{_{\rm SCL}}\!(t)}{n_{_{\rm FIL}}}\right]^{1/2}\;\,=\;\,\frac{c_{_{\rm II}}}{c_{_{\rm SCL}}}\left[\frac{n_{_{\rm IF}}(t)}{n_{_{\rm FIL}}}\right]^{1/2}\,,
\end{eqnarray}
Eq. \ref{EQN:dSdt.1} becomes 
\begin{eqnarray}\label{EQN:dSdt.2}
\frac{dS}{dt}&\approx&c_{_{\rm SCL}}\,\chi(t)\,.
\end{eqnarray}

Combining Eqs. \ref{EQN:vSCL.1} and \ref{EQN:dLdt.1},
\begin{eqnarray}\label{EQN:dLdt.2}
\frac{dL}{dt}\!&\!\!\approx\!\!&\!\frac{dS}{dt}\,-\,\frac{c_{_{\rm SCL}}^2}{dS/dt}\,+\,\frac{c_{_{\rm SCL}}^2}{2\,c_{_{\rm II}}}\;\approx\;c_{_{\rm SCL}}\left\{\chi(t)-\frac{1}{\chi(t)}\right\},\hspace{0.5cm}
\end{eqnarray}
where, to obtain the final expression, we have dropped the third term ($c_{_{\rm FIL}}^2/2c_{_{\rm II}}$) in the middle expression; we justify this in \S\ref{SUBSEC:checkshockvelocity}, by showing that, apart from a very short period just after the O star switches on, $dS/dt\ll 2c_{_{\rm II}}$.

Finally, from Eq. \ref{EQN:W.1} we have
\begin{eqnarray}\label{EQN:dWdt.1}
\frac{dW}{dt}&=&\frac{dS}{dt}-\frac{dL}{dt}\;\,\approx\;\,\frac{c_{_{\rm SCL}}}{\chi(t)}\,;
\end{eqnarray}
from Eq. \ref{EQN:shock.1},
\begin{eqnarray}\label{EQN:nSCL.1}
n_{_{\rm SCL}}\!(t)&\approx&n_{_{\rm FIL}}\,\chi^2(t)\,;
\end{eqnarray}
and from Eq. \ref{EQN:cFIL.1}  (see \S\ref{SEC:Model}),
\begin{eqnarray}\label{EQN:cSCL.1}
c_{_{\rm SCL}}&=&\left[\pi Gn_{_{\rm FIL}}{\bar m}_{_{\rm H_2}}\right]^{1/2}\,R_{_{\rm FIL}}\\\label{EQN:cSCL.2}
&\rightarrow&0.615\,{\rm km\,s^{-1}}\;R_{_{0.2}}\;n_{_4}^{1/2}\,.
\end{eqnarray}

\section{Advance of the ionisation front}\label{SEC:IFlocation}

We now introduce the parameter,
\begin{eqnarray}\label{EQN:K1.1}
{\cal K}_{_{\rm O}}&=&\left[\frac{1}{\pi G{\bar m}_{_{\rm H_2}}}\right]^{1/2}\frac{c_{_{\rm II}}}{R_{_{\rm FIL}}n_{_{\rm FIL}}}\,\left[\frac{3\,\dot{\cal N}_{_{\rm LyC}}}{4\,\pi\,\alpha_{_{\rm B}}\,R_{_{\rm FIL}}^3}\right]^{1/4}\hspace{0.5cm}\\
&\rightarrow&19\;\;R_{_{0.2}}^{-7/4}\;n_{_4}^{-1}\;\dot{\cal N}_{_{\rm 49}}^{1/4}
\end{eqnarray}
(which measures the speed with which the O star erodes the filament) and the dimensionless length and time variables,
\begin{eqnarray}\label{EQN:lambda.1} 
\lambda(\tau)\!\!&\!\!=\!\!&\!\!\frac{L}{R_{_{\rm FIL}}}\;\rightarrow\;\frac{L}{0.2\,{\rm pc}\;R_{_{0.2}}}\,,\\\label{EQN:tau.1}
\tau\!\!&\!\!=\!\!&\!\!\frac{c_{_{\!\rm SCL}}\,t}{R_{_{\rm FIL}}}\;=\;\left[2\pi Gn_{_{\rm FIL}}{\bar m}_{_{\rm H_2}}\right]^{1/2}t\;\rightarrow\;\frac{t}{0.22\,{\rm Myr}\;n_{_4}^{-1/2}}.\hspace{0.6cm}
\end{eqnarray}
$n_{_{\rm IF}}(t)$ can then be eliminated between Eqs. \ref{EQN:IonBal.2}, \ref{EQN:chi.1}, \ref{EQN:dLdt.2}, \ref{EQN:cSCL.1} and \ref{EQN:K1.1} to give
\begin{eqnarray}\label{EQN:dlambdadtau.1}
\frac{d\lambda}{d\tau}&\approx&\chi(\tau)\,-\,\frac{1}{\chi(\tau)}\,,\\\label{EQN:chi.2}
\chi(\tau)&=&{\cal K}_{_{\rm O}}\,\left\{\lambda^2(\tau)-\lambda(\tau)+1-\left[1+\lambda(\tau)\right]^{-1}\right\}^{-1/4},\hspace{0.8cm}
\end{eqnarray}
Eq. \ref{EQN:dlambdadtau.1} is the equation of motion for the IF, and must be solved numerically.

\begin{figure}
\vspace{-1.20cm}\hspace{-0.90cm}\includegraphics[angle=270.0,width=1.27\columnwidth]{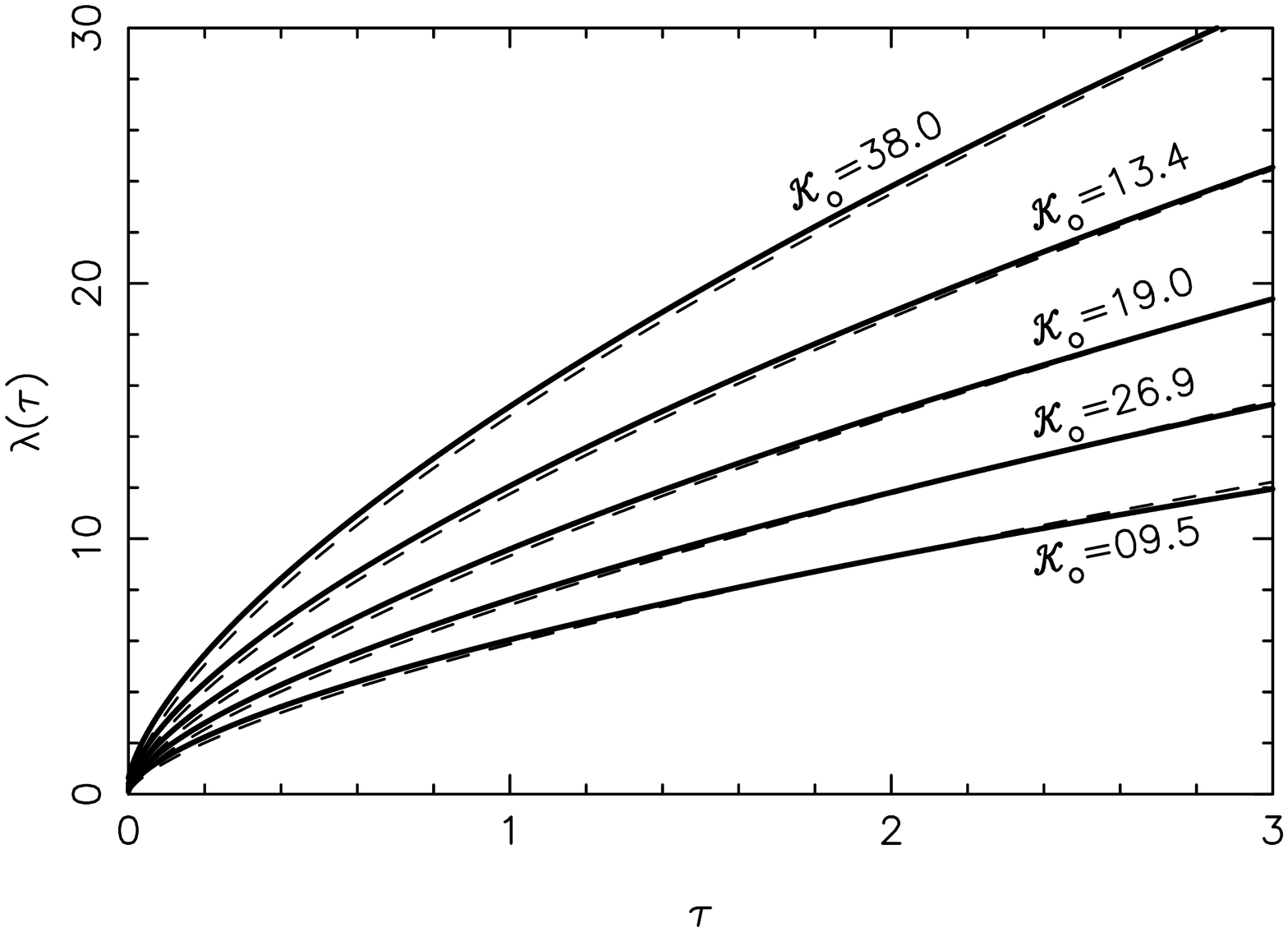}
\vspace{-0.5cm}\caption{The full curves show numerical integrations of Eq. \ref{EQN:dlambdadtau.1} for ${\cal K}_{_{\rm O}}=09.5,\;13.4,\;19.0,\;26.9\;{\rm and}\;38.0$; this gives the position of the IF as a function of time, in dimensionless units (Eqs. \ref{EQN:lambda.1} and \ref{EQN:tau.1}). The dashed curves show the approximate asymptotic solution (Eq. \ref{EQN:lambda.2}) for the same values of ${\cal K}_{_{\rm O}}$.}
\label{FIG:lambda(tau)}
\end{figure}

We start the integration of Eq. \ref{EQN:dlambdadtau.1} with $L$ equal to the Str{\o}mgren radius, i.e.
\begin{eqnarray}\nonumber
L(0)\;\;=\;\;R_{_{\rm Str}}&=&\left[\frac{3\,\dot{\cal N}_{_{\rm LyC}}}{4\,\pi\,\alpha_{_{\rm B}}\,n_{_{\rm FIL}}^2}\right]^{1/3}\\\label{EQN:L0.1}
&\rightarrow&0.144\,{\rm pc}\;\;n_{_4}^{-2/3}\;\dot{\cal N}_{_{49}}^{1/3}\,,\hspace{0.5cm}
\end{eqnarray}
hence
\begin{eqnarray}\label{EQN:lambda0.1}
\lambda(0)&\rightarrow&0.722\;\;R_{_{0.2}}^{-1}\;n_{_4}^{-2/3}\;\dot{\cal N}_{_{\rm 49}}^{1/3}\,.
\end{eqnarray}
Accurate numerical solutions for $\lambda(\tau)$, obtained by integrating Eqs. \ref{EQN:dlambdadtau.1} and \ref{EQN:chi.2}, with initial condition given by Eq. \ref{EQN:lambda0.1}, are plotted with full lines on Fig. \ref{FIG:lambda(tau)}, for five representative values of ${\cal K}_{_{\rm O}}=9.5,\,13.4,\,19.0,\,26.9,\,38.0$.

We can also seek an approximate asymptotic solution by taking the leading terms in Eqs. \ref{EQN:dlambdadtau.1} and \ref{EQN:chi.2}, to obtain
\begin{eqnarray}\label{EQN:dlambdadtau.2}
\frac{d\lambda}{d\tau}&\!\sim\!&\chi(\tau)\;\;\sim\;\;{\cal K}_{_{\rm O}}\,\lambda^{-1/2}(\tau)\,,
\end{eqnarray}
and hence
\begin{eqnarray}\label{EQN:lambda.2}
\lambda(\tau)&\!\sim\!&\left[\frac{3\,{\cal K}_{_{\rm O}}\,\tau}{2}\right]^{2/3},\hspace{1.2cm}
\end{eqnarray}
and
\begin{eqnarray}\label{EQN:chi.3}
\chi(\tau)\!&\!\!\sim\!\!&\!\left[\frac{2\,{\cal K}_{_{\rm O}}^2}{3\,\tau}\right]^{1/3}\\
\!&\!\!\rightarrow\!\!&\!3.77\;\;R_{_{0.2}}^{-7/6}\;n_{_4}^{-5/6}\;\dot{\cal N}_{_{49}}^{1/6}\;\left[t/{\rm Myr}\right]^{-1/3}\!.\hspace{0.5cm}
\end{eqnarray}
The approximate asymptotic solution (Eq. \ref{EQN:lambda.2}) is plotted with dashed lines on Fig. \ref{FIG:lambda(tau)}, for the same five values of ${\cal K}_{_{\rm O}}$. There is very close agreement between the accurate numerical solution and the approximate asymptotic solution. Since the latter is analytic, we shall use it in the sequel to estimate other properties of the H{\sc ii}R and the SCL.

Converting Eq. \ref{EQN:lambda.2} back to physical variables gives
\begin{eqnarray}\label{EQN:L.2}
L(t)\!\!&\!\!\sim\!\!&\!\!5.2\,{\rm pc}\;R_{_{0.2}}^{-1/6}\,n_{_4}^{-1/3}\,\dot{\cal N}_{_{49}}^{1/6}\,\left[t/{\rm Myr}\right]^{2/3}\!,\\\label{EQN:dLdt.3}
\frac{dL}{dt}\!\!&\!\!\sim\!\!&\!\!3.3\,{\rm km\,s^{-1}}\;R_{_{0.2}}^{-1/6}\,n_{_4}^{-1/3}\,\dot{\cal N}_{_{49}}^{1/6}\,\left[t/{\rm Myr}\right]^{-1/3}\!,
\end{eqnarray}
and
\begin{eqnarray}
t(L)\!\!&\!\!\sim\!\!&\!\!0.085\,{\rm Myr}\;R_{_{0.2}}^{1/4}\,n_{_4}^{1/2}\,\dot{\cal N}_{_{49}}^{-1/4}\,\left[L/{\rm pc}\right]^{3/2}\,.\hspace{1.0cm}
\end{eqnarray}

\section{Dynamics of the HII region}\label{HIIRdynamics}

From Eq.\ref{EQN:chi.1}, the density of ionised gas at the IF is
\begin{eqnarray}\label{EQN:nIF.1}
n_{_{\rm IF}}(t)&\approx&n_{_{\rm FIL}}\left[\frac{c_{_{\rm SCL}}\,\chi(t)}{c_{_{\rm II}}}\right]^2\\\label{EQN:nIF.2}
&\sim&240\,{\rm cm^{-3}}\;\;R_{_{0.2}}^{-1/3}\;n_{_4}^{1/3}\;\dot{\cal N}_{_{49}}^{1/3}\;\left[t/{\rm Myr}\right]^{-2/3}.\hspace{0.9cm}
\end{eqnarray}

Strictly speaking, the expressions and estimates derived heretofore (Eqs. \ref{EQN:IonBal.1} through \ref{EQN:nIF.2}) pertain only to the gas on the $z$ axis. The flux of ionising radiation incident on other parts of the IF decreases with distance from the $z$ axis, largely because it arrives at increasingly oblique angles. To take account of this we invoke the second of the {\it Ans{\"a}tze} defined in \S\ref{SEC:Model}. We assume that the expressions and estimates derived heretofore obtain everywhere, not just on the $z$ axis, and to compensate for this we limit the area of the IF to the cross-sectional area of the filament, $\pi R_{_{\rm FIL}}^2$. The IF is therefore one quarter of a sphere, i.e. tetartospherical. On Fig. \ref{FIG:Cartoon_FiF} the outermost flow arrows correspond to gas flowing off the edge of the IF.

The net rate at which ionised gas flows off the IF is
\begin{eqnarray}
dM_{_{\rm II}}/dt\!\!&\!\!\sim\!\!&\!\!\pi\,R_{_{\rm FIL}}^2\,n_{_{\rm IF}}(t)\,\bar{m}_{_{\rm H}}\,c_{_{\rm II}}\\
\!\!&\!\!\sim\!\!&\!\!16\,{\rm M_{_\odot}\,Myr^{-1}}\,R_{_{0.2}}^{5/3}\,n_{_4}^{1/3}\,\dot{\cal N}_{_{49}}^{1/3}
\left[t/{\rm Myr}\right]^{-2/3}\hspace{0.5cm}
\end{eqnarray}
The total mass boiled off the IF is therefore
\begin{eqnarray}
M_{_{\rm II}}&\sim&48\,{\rm M_{_\odot}}\;\;R_{_{0.2}}^{5/3}\;n_{_4}^{1/3}\;\dot{\cal N}_{_{49}}^{1/3}\;\left[t/{\rm Myr}\right]^{1/3}\hspace{1.00cm}
\end{eqnarray}

The fraction of ionising radiation expended maintaining the flow of ionised gas off the end of the filament, and out to infinity, is given by
\begin{eqnarray}\label{EQN:fLyC.1}
f_{_{\rm LyC}}\!&\!\sim\!&\!\int\limits_{r=R_{_{\rm FIL}}}^{r=\infty}\,\alpha_{_{\rm B}}\,n_{_{\rm IF}}^2\left[\frac{r}{R_{_{\rm FIL}}}\right]^{-4}\,\pi\,r^2\,dr\;\sim\;\frac{3}{4}\left[\frac{L}{R_{_{\rm FIL}}}\right]^{-2}\hspace{0.6cm}\\
\!&\!\sim\!&\!0.030\;R_{_{0.2}}^2\,\left[L/{\rm pc}\right]^{-2}\\
\!&\!\sim\!&\!0.11\;R_{_{0.2}}^{7/3}\;n_{_4}^{2/3}\;\dot{\cal N}_{_{49}}^{-1/3}\,\left[t/{\rm Myr}\right]^{-4/3}.
\end{eqnarray}
For our fiducial configuration parameters, $f_{_{\rm LyC}}$ quickly becomes very small; a large fraction of the ionising radiation escapes.

\section{Growth of the shock-compressed layer}\label{SEC:SCLgrowth}

\begin{figure}
\vspace{-1.20cm}\hspace{-0.90cm}\includegraphics[angle=270.0,width=1.27\columnwidth]{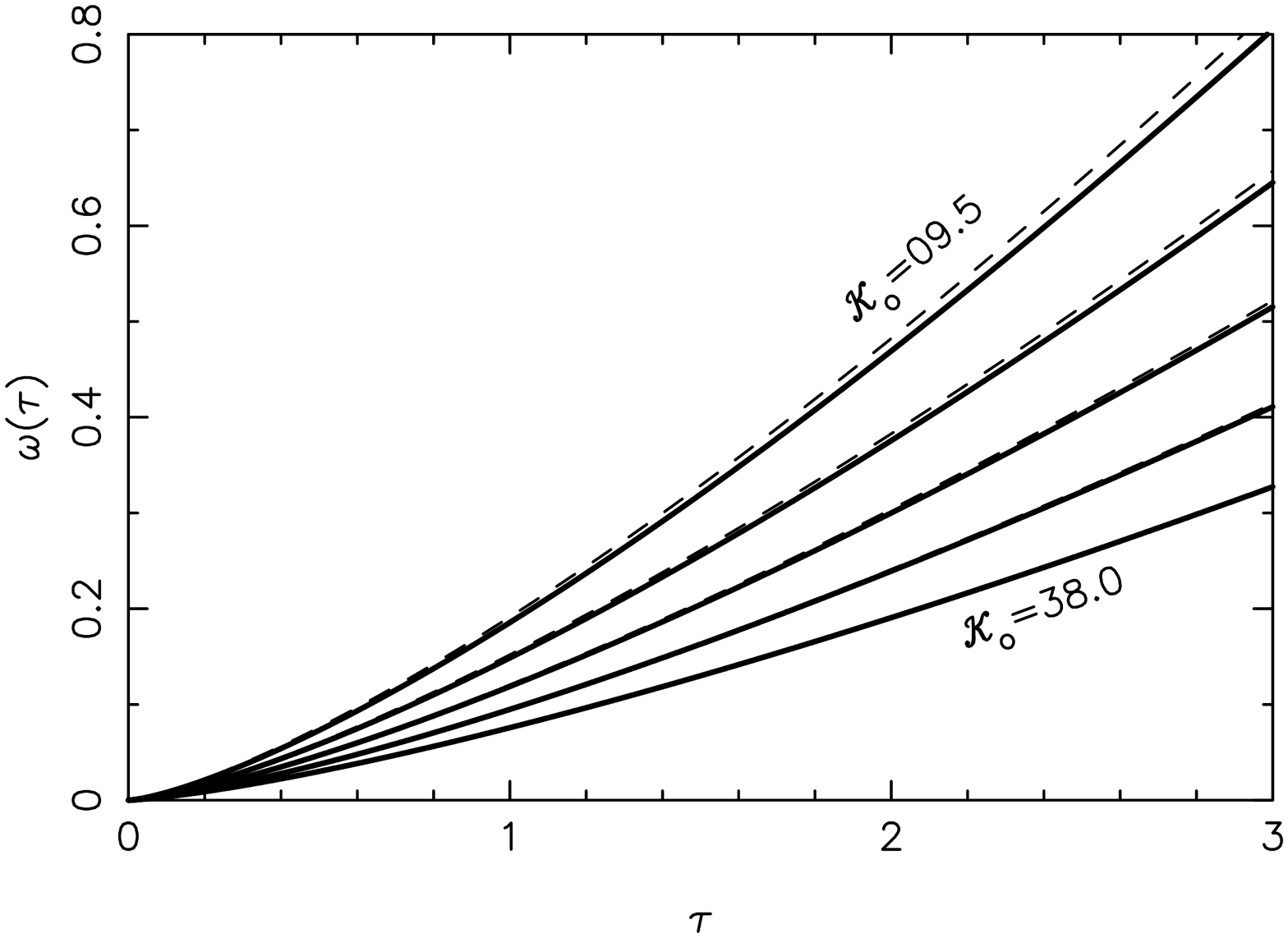}
\vspace{-0.5cm}\caption{The full curves show numerical integrations of Eq. \ref{EQN:domegadtau.1} for ${\cal K}_{_{\rm O}}=09.5,\;13.4,\;19.0,\;26.9\;{\rm and}\;38.0$; this gives the thickness of the SCL as a function of time, in dimensionless units. The dashed curves show the approximate asymptotic solution (Eq. \ref{EQN:omegatau.1}) for the same values of ${\cal K}_{_{\rm O}}$.}
\label{FIG:omega(tau)}
\end{figure}

The thickness of the SCL grows at a rate given by Eq. \ref{EQN:dWdt.1}, and hence in terms of dimensionless variables,
\begin{eqnarray}\label{EQN:domegadtau.1}
\frac{d\omega}{d\tau}&\approx&\frac{1}{\chi(\tau)}
\end{eqnarray}
Accurate numerical solutions for $\omega(\tau)$ can be obtained by integrating Eq. \ref{EQN:domegadtau.1}, with $\chi$ from Eq. \ref{EQN:chi.2} and the initial condition $\omega(0)=0$. Accurate numerical solutions for $\omega(\tau)$, with ${\cal K}_{_{\rm O}}=9.5,\,13.4,\,19.0,\,26.9,\,38.0$, are plotted with full lines on Fig.~\ref{FIG:omega(tau)}.

In the limit that $\lambda\gg 1$, we can substitute from Eq. \ref{EQN:chi.3} in Eq. \ref{EQN:domegadtau.1} to obtain
\begin{eqnarray}
\frac{d\omega}{d\tau}&\sim&\left[\frac{3\,\tau}{2\,{\cal K}_{_{\rm O}}^2}\right]^{1/3}\,,
\end{eqnarray}
and hence an approximate asymptotic solution,
\begin{eqnarray}\label{EQN:omegatau.1}
\omega(\tau)&\sim&\left[\frac{81\,\tau^4}{128\,{\cal K}_{_{\rm O}}^2}\right]^{1/3}\,.
\end{eqnarray}
The approximate asymptotic solution (Eq. \ref{EQN:omegatau.1}) is plotted with dashed lines on Fig. \ref{FIG:omega(tau)}, for the same five values of ${\cal K}_{_{\rm O}}$. Again there is very good agreement between the accurate numerical solution and the approximate asymptotic solution, and we adopt the latter because it is analytic. Converting back to physical variables, Eq. \ref{EQN:omegatau.1} gives
\begin{eqnarray}\label{EQN:W.2}
W(t)\!\!&\!\sim\!&\!\!0.18\,{\rm pc}\;R_{_{0.2}}^{13/6}\,n_{_4}^{4/3}\,\dot{\cal N}_{_{49}}^{-1/6}\left[t/{\rm Myr}\right]^{4/3}.\hspace{0.5cm}
\end{eqnarray}
Combining Eqs. \ref{EQN:L.2} and \ref{EQN:W.2}, we obtain
\begin{eqnarray}
W(t)/L(t)&\sim&0.032\;R_{_{0.2}}^{7/3}\,n_{_4}^{5/3}\,\dot{\cal N}_{_{49}}^{-1/3}\left[t/{\rm Myr}\right]^{4/3}\!.\hspace{0.6cm}
\end{eqnarray}
Thus, unless the radius and/or density of the undisturbed filament are very large, and/or the erosion has been ongoing for a very long time, $\;W(t)\!\ll\!L(t)$, and so to a first approximation we can set $S(t)\sim L(t)$.

Most of the swept-up mass is in the SCL -- rather than in the ionised outflow -- so the mass of the SCL is
\begin{eqnarray}
M_{_{\rm SCL}}(t)\!\!&\!\!\sim\!\!&\!\!\pi\,R_{_{\rm FIL}}^2\,L(t)\,n_{_{\rm FIL}}\,{\bar m}_{_{\rm H_2}}\\
\!\!&\!\!\sim\!\!&\!\!450\,{\rm M_{_\odot}}\;\;R_{_{0.2}}^{11/6}\;n_{_4}^{2/3}\;\dot{\cal N}_{_{49}}^{1/6}\;\left[t/{\rm Myr}\right]^{2/3},\hspace{1.0cm}
\end{eqnarray}
its volume-density is
\begin{eqnarray}
n_{_{\rm SCL}}(t)\!\!&\!\!\!\sim\!\!\!&\!\!n_{_{\rm FIL}}\,L(t)/W(t)\\
\!\!&\!\!\!\sim\!\!\!&\!\!2.8\!\times\! 10^5{\rm cm^{-3}}R_{_{0.2}}^{-7/3}n_{_4}^{-5/3}\dot{\cal N}_{_{49}}^{1/3}\!\left[t/{\rm Myr}\right]^{-2/3},\hspace{0.5cm}
\end{eqnarray}
and its surface-density (looking from the O star) is
\begin{eqnarray}\label{EQN:SigmaSCL.1}
\Sigma_{_{\rm SCL}}(t)\!\!&\!\!\sim\!\!&\!\!W(t)\,n_{_{\rm SCL}}\!(t)\,\bar{m}_{_{\rm H_2}}\\\label{EQN:SigmaSCL.2}
\!\!&\!\!\sim\!\!&\!\!3700\,{\rm M_{_\odot}\,pc^{-2}}\,R_{_{0.2}}^{-1/6}n_{_4}^{2/3}\dot{\cal N}_{_{49}}^{1/6}\!\left[t/{\rm Myr}\right]^{2/3}.\hspace{0.7cm}
\end{eqnarray}
The surface-density quickly exceeds the notional threshold for efficient star formation, $\Sigma_{_{\rm SF}}\sim 160\,{\rm M_{_\odot}\,pc^{-2}}$, corresponding to visual extinction $A_{_{\rm V}}\sim 8\,{\rm mag}$ \citep[e.g.][]{AndrePhetal2010, LadaCetal2010, KonyvesVetal2013, KonyvesV2015}. Therefore we might expect a second generation of star formation to be triggered in the SCL, unless -- or even if -- there is already star formation ongoing there.

Combining Eqs. \ref{EQN:vSCL.1} and \ref{EQN:dSdt.2}, the velocity of the gas in the SCL (away from the O star) is
\begin{eqnarray}
v_{_{\rm SCL}}\!(t)&\sim&c_{_{\rm SCL}}\,\chi(t)\\
&\sim&1.9\,{\rm km/s}\,R_{_{0.2}}^{-1/6}n_{_4}^{-1/3}\;\dot{\cal N}_{_{49}}^{1/6}\;\left[t/{\rm Myr}\right]^{-1/3}.\hspace{0.9cm}
\end{eqnarray}

\section{Approximations and assumptions}\label{SEC:Approximations}

\subsection{Slow advance of the IF}\label{SUBSEC:checkslowIFadvance}

In deriving Eq. \ref{EQN:RecombRate} for the recombination rate, we have assumed that the timescale on which newly ionised gas disperses is much shorter than the timescale on which the IF advances (i.e. Eq. \ref{INEQ:subsonicIFSF}), and therefore that the dependence on $r$ (distance from the CC) can be separated from the dependence on $t$ (time). Substituting from the analytic solutions for $L(t)$ (Eq. \ref{EQN:L.2}) and $dL/dt$ (Eq. \ref{EQN:dLdt.3}), Eq. \ref{INEQ:subsonicIFSF} reduces to
\begin{eqnarray}
t&\gg&0.009\,{\rm Myr}\,R_{_{0.2}}\,.
\end{eqnarray}
In other words, there is a very short period when the O star first switches on and this assumption is invalid, but thereafter the assumption becomes increasingly accurate.

\subsection{Fraction of ionising radiation reaching the IF}\label{SUBSEC:checkIFfraction}

In deriving Eq. \ref{EQN:IonBal.2}, we have assumed that all the ionising radiation is expended balancing recombination, i.e. we have neglected the ionising photons that reach the IF and ionise new material. Of the ionising photons emitted in the direction of the exposed end of the filament, the fraction that does actually reach the IF and ionise new material is
\begin{eqnarray}
f_{_{\rm IF}}&\approx&\frac{4\,\pi\,L^2(t)\,n_{_{\rm IF}}(t)\,c_{_{\rm II}}}{\dot{\cal N}_{_{\rm LyC}}}\\
&\sim&0.11\,R_{_{0.2}}^{-2/3}n_{_4}^{-1/3}\dot{\cal N}_{_{49}}^{-1/3}\!\left[t/{\rm Myr}\right]^{2/3}\hspace{0.9cm}\\
&\sim&0.02\;\;R_{_{0.2}}^{-1/2}\;\dot{\cal N}_{_{49}}^{-1/2}\;\left[L/{\rm pc}\right],\hspace{0.9cm}
\end{eqnarray}
Evidently this fraction is low -- and therefore we do not incur a large error by ignoring it in the equation of ionisation balance (Eq. \ref{EQN:IonBal.2}) -- unless we are considering the very late stages of a very weak ionising source, dispersing a very thin and rarefied filament.

\subsection{Speed of advance of the SF}\label{SUBSEC:checkshockvelocity}

In deriving Eq. \ref{EQN:shock.1} for the density in the SCL, we have assumed that the SF advance supersonically relative to the neutral gas ahead of it, i.e.
\begin{eqnarray}\label{EQN:shockcomp}
dS/dt\sim dL/dt \gg c_{_{\rm FIL}}.
\end{eqnarray}
Substituting for $dL/dt$ from Eq. \ref{EQN:dLdt.3}, and for $c_{_{\rm FIL}}$ from Eq. \ref{EQN:cFIL.2}, Eq. \ref{EQN:shockcomp} reduces to
\begin{eqnarray}
t&\ll&450\,{\rm Myr}\;R_{_{0.2}}^{-7/2}n_{_4}^{-5/2}\dot{\cal N}_{_{49}}^{1/2}.
\end{eqnarray}
Unless the filament is very fat (large $R_{_{\rm FIL}}$) and/or very dense (large $n_{_{\rm FIL}}$) and/or the ionising output of the `O star' is very weak (small $\dot{\cal N}_{_{\rm LyC}}$), this condition is easily satisfied.

To obtain the final expression in Eq. \ref{EQN:dLdt.2}, we have assumed that the SF advances much more slowly than $2c_{_{\rm II}}$, i.e.
\begin{eqnarray}\label{EQN:dSdt.5}
dS/dt\sim dL/dt \ll 2c_{_{\rm II}}.
\end{eqnarray}
Substituting for $dL/dt$ from Eq. \ref{EQN:dLdt.3}, Eq. \ref{EQN:dSdt.5} reduces to
\begin{eqnarray}
t&\gg&0.0013\,{\rm Myr}\;R_{_{0.2}}^{-1/2}n_{_4}^{-1}\dot{\cal N}_{_{49}}^{1/2}.
\end{eqnarray}
Unless the filament is very thin (small $R_{_{\rm FIL}}$) and/or very diffuse (small $n_{_{\rm FIL}}$) and/or the ionising output of the O star is very strong (large $\dot{\cal N}_{_{\rm LyC}}$), this condition is easily satisfied.

\subsection{Radiation pressure and stellar wind from the O star}\label{SEC:Assumptions}

In addition to the assumptions inherent in the two {\it Ans{\"a}tze} of Section 2, the model ignores the fact that, as the outward flow of ionised gas approaches the O star, it will encounter radiation pressure from the star, and probably also a wind, both of which will deflect the flow away from the $z$ axis. However, for the purpose of estimating the rate at which the ionisation front erodes the end of the filament, these factors are of secondary importance. This is because, once $L(t)>R_{_{\rm FIL}}$, attenuation of the ionising flux reaching the end of the filament is dominated by the region close to the IF, as evidenced by the form of the integral in Eq. \ref{EQN:IonBal.1}.

\subsection{O star formed in the middle of a filament}\label{SUBSEC:Ostarinthemiddle}

The situation is a little more complicated if the O star forms in the middle of a filament, so that it is eroding the filament on two opposing fronts. In this case, as the flow of ionised gas approaches the O star from one side it encounters a flow coming from the other side. Consequently there is a contact discontinuity, or more probably a turbulent mixing-layer, between the two flows, and they are deflected even more effectively away from the $z$ axis. However, this is again of secondary importance because the important region is the immediate vicinity of the IF where the gas is rapidly diverging.

\subsection{Magnetic field}

We have neglected the likelihood that there is a significant magnetic field. Observations suggest that dominant filaments tend to be oriented perpendicular to the local large-scale magnetic field \citep[e.g.][]{SolerJetal2017, Arzoumanetal2020}. If this is the case, the magnetic field will have two main effects. First, the magnetic pressure will weaken the strength of the SF and decelerate the advance of the IF into the filament. Second, the flow of ionised gas off the IF will follow/advect the field lines, and is therefore likely to disperse more quickly; this will increase the flux of ionising radiation reaching the IF, thereby accelerating the advance of the IF into the filament. These two effects act in opposite senses, but probably the first one dominates. The net effect can only be evaluated properly by specifying the strength of the magnetic field and performing a full MHD simulation.

\subsection{Displacement of the ionising star}

The ionising star, or stars, will be born with (and/or acquire through dynamical interactions with other stars) a finite velocity. Consequently it will become displaced from the $z$ axis defining the spine of the filament. On simple geometric grounds, this will only seriously compromise the model if the velocity is greater than the speed at which the IF advances (Eq. \ref{EQN:dLdt.3}). Even then, if the velocity is mostly directed parallel, or anti-parallel, to the $z$ axis, the model can easily be adjusted to account for this.

\subsection{Continuing accretion flow onto the filament}\label{SEC:inflow}

Finally, we have neglected the possibility that there is a continuing accretion flow onto the filament -- i.e. a continuation of the convergent flow that created the filament in the first place -- and that this may act to trap the ionised gas boiling off the end of the filament. A proper evaluation of the effect that such an accretion flow might have on the dynamics of the ionised gas lies outside the scope of this paper, but in the next section (\S\ref{SEC:trappedHIIregion}) we estimate the circumstances under which it is likely to be an important effect. We conclude that it is only an issue if there is a very dense and rapid continuing accretion flow onto a filament with large $R_{_{\rm FIL}}$ and/or large $n_{_{\rm FIL}}$, and/or a weak ionising source (low $\dot{\cal N}_{_{\rm LyC}}$).

\section{Trapped HII Region}\label{SEC:trappedHIIregion}

In this penultimate section we assess whether continuing accretion onto the filament might affect the dynamics of the ionised gas flowing off the end of the filament.

First, the total pressure associated with the ionised gas flowing off the IF is given by
\begin{eqnarray}
\frac{P_{_{\rm IF}}(t)}{k_{_{\rm B}}}\!\!&\!\!\sim\!\!&\!\!\frac{2n_{_{\rm IF}}(t){\bar m}_{_{\rm H}}c_{_{\rm II}}^2}{k_{_{\rm B}}}\\
\!\!&\!\!\sim\!\!&\!\!29\!\times\!10^5{\rm cm^{-3}K}\,R_{_{0.2}}^{-1/3}n_{_4}^{1/3}\dot{\cal N}_{_{49}}^{1/3}\!\left[t/{\rm Myr}\right]^{-2/3}\!.\hspace{0.4cm}
\end{eqnarray}
For typical values of the configuration parameters, this is quite a large pressure, by interstellar standards. For example, the critical surface-density threshold for star formation, $\Sigma_{_{\rm SF}}\sim 160\,{\rm M_{_\odot}\,pc^{-2}}$ corresponds to a critical pressure $P_{_{\rm SF}}/k_{_{\rm B}} \sim G\Sigma_{_{\rm SF}}^2/k_{_{\rm B}} \sim 5\times 10^5\,{\rm cm^{-3}\,K}$. Therefore the critical region near the IF, where in our model most of the ionising photons from the O star are expended, should be maintained unless the ram-pressure of the continuing inflow is very large. (Nonetheless, we should be mindful that only part of the momentum carried by the flow of ionised gas is directed away from the $z$ axis. Furthermore the flux of momentum carried by this flow decreases approximately as $r^{-2}$, where $r$ is distance from the CC.)

Second, the pressure in the undisturbed filament is given by
\begin{eqnarray}
\frac{P_{_{\rm FIL}}}{k_{_{\rm B}}}&\sim&\frac{n_{_{\rm FIL}}\bar{m}_{_{\rm H_2}}c_{_{\rm FIL}}^2}{k_{_{\rm B}}}\\
&\rightarrow&6.5\times 10^5\,{\rm cm^{-3}\,K}\;\;R_{_{0.2}}^2\;n_{_4}^2\,,
\end{eqnarray}
and since the filament is self-gravitating, the ram-pressure on its surface, due to the continuing accretion flow, must be less than, or at most comparable with, $P_{_{\rm FIL}}$. Moreover, as we move away from the filament the ram-pressure of the accretion flow is likely to decrease, either because it is petering out (there is a finite supply of mass), or because the density goes down (it is a converging flow), or because the inward velocity goes down (the inward velocity is accelerated somewhat by the gravitational field of the filament).

For the purpose of making a simple and highly conservative estimate of the effect of the continuing accretion flow, we assume (i) that the accretion flow is not petering out and the ram pressure it delivers at the surface of the filament is comparable with $P_{_{\rm FIL}}$; (ii) that the inward velocity is constant (for inward velocities $\ga 1\,{\rm km\,s^{-1}}$, the acceleration due to the gravitational field of the filament, between a distance $r_{_{\!\perp}}\sim 2\,{\rm pc}$ from the $z$ axis and $r_{_{\!\perp}}= R_{_{\rm FIL}}$, is small for all reasonable values of $R_{_{\rm FIL}}$); and (iii) that the density declines as $r_{_{\!\perp}}^{-1}$ (so the rate of inflow across cylindrical surfaces of radius $r_{_{\!\perp}}$, centred on the $z$ axis, is independent of $r_{_{\!\perp}}$). It follows that the ram-pressure of the accretion flow is
\begin{eqnarray}
\frac{P_{_{\rm ACC}}(r_{_{\!\perp}})}{k_{_{\rm B}}}&\approx&\frac{P_{_{\rm FIL}}R_{_{\rm FIL}}}{k_{_{\rm B}}r_{_{\!\perp}}}\\
&\rightarrow&1.3\!\times\!10^5\,{\rm cm^{-3}\,K}\;R_{_{0.2}}^3\,n_{_4}^2\left[r_{_{\!\perp}}/{\rm pc}\right]^{-1}.\hspace{0.8cm}
\end{eqnarray}
This estimate of $P_{_{\rm ACC}}$ is conservative in the sense that the three preceding assumptions (i, ii and iii) all maximise the ram-pressure of the accretion flow, and hence its ability to trap the H{\sc ii} region.

If the accretion flow is effective in trapping the H{\sc ii} region, the H{\sc ii} region adopts an approximately uniform density, ${\bar n}_{_{\rm HIIR}}$ throughout most of its volume. (The divergent velocity field of the gas flowing off the end of the filament is confined to a very small region and terminates in a shock where this gas merges with the uniform-density H{\sc ii}.) The mean (direction-averaged) radius of the H{\sc ii} region, ${\bar R}_{_{\rm HIIR}}$, is therefore given by
\begin{eqnarray}
\frac{4\pi\alpha_{_{\rm B}}{\bar n}_{_{\rm HIIR}}^2{\bar R}_{_{\rm HIIR}}^3}{3}&\approx&\dot{\cal N}_{_{\rm LyC}}\,,
\end{eqnarray} 
and its mean pressure is
\begin{eqnarray}
\frac{{\bar P}_{_{\rm HIIR}}(\bar{R}_{_{\rm HIIR}})}{k_{_{\rm B}}}\!&\!\approx\!&\!\left[\frac{3\dot{\cal N}_{_{\rm LyC}}}{4\pi\alpha_{_{\rm B}}{\bar R}_{_{\rm HIIR}}^3}\right]^{1/2}\frac{{\bar m}_{_{\rm H}}c_{_{\rm II}}^2}{k_{_{\rm B}}}\\
\!&\!\rightarrow\!&\!210\times 10^5\,{\rm cm^{-3}\,K}\;\;\dot{\cal N}_{_{49}}^{1/2}\,\left[{\bar R}_{_{\rm HIIR}}/{\rm pc}\right]^{-3/2}.\hspace{0.8cm}
\end{eqnarray}

If the H{\sc ii} region is confined by the ram-pressure of the accretion flow, then $P_{_{\rm ACC}}\!({\bar R}_{_{\rm HIIR}})\approx{\bar P}_{_{\rm HIIR}}\!(\bar{R}_{_{\rm HIIR}})$ and the mean radius and mean density of the H{\sc ii} region are
\begin{eqnarray}\label{EQN:RHIIR.1}
\bar{R}_{_{\rm HIIR}}&\approx&\frac{3\,\dot{\cal N}_{_{\rm LyC}}\,c_{_{\rm II}}^4}{16\,\pi^3\,\alpha_{_{\rm B}}\,G^2\,R_{_{\rm FIL}}^6\,n_{_{\rm FIL}}^4\,\bar{m}_{_{\rm H}}^2}\\\label{EQN:RHIIR.2}
&\rightarrow&2.65\times 10^4\,{\rm pc}\;\;R_{_{0.2}}^{-6}\;n_{_4}^{-4}\;\dot{\cal N}_{_{49}},\\\label{EQN:nHIIR.1}
\bar{n}_{_{\rm HIIR}}&\approx&\frac{32\,\pi^4\,\alpha_{_{\rm B}}\,G^3\,R_{_{\rm FIL}}^9\,n_{_{\rm FIL}}^6\,\bar{m}_{_{\rm H}}^3}{3\,\dot{\cal N}_{_{\rm LyC}}\,c_{_{\rm II}}^6}\\\label{EQN:nHIIR.2}
&\rightarrow&1.27\times 10^{-4}\,{\rm cm^{-3}}\;\;R_{_{0.2}}^9\;n_{_4}^6\;\dot{\cal N}_{_{49}}^{-1}\,.
\end{eqnarray}
With the fiducial values we have chosen for the configuration parameters, Eqs. \ref{EQN:RHIIR.2} and \ref{EQN:nHIIR.2} suggest that the H{\sc ii} region is very unlikely to be trapped by a continuing inflow (this would require continuing inflow at radii well beyond $\bar{R}_{_{\rm HIIR}}\sim 26.5\,{\rm kpc}$ from the filament spine, which is seriously implausible), and indeed this is probably the case for most filaments hosting O stars. 

However, the dependence on the configuration parameters is very strong, and continuing inflow onto a relatively broad, relatively dense filament, with a relatively weak ionising source, could lead to a trapped H{\sc ii} region. For example, if $R_{_{\rm FIL}}\!\sim\!0.7\,{\rm pc}$ ($R_{_{0.2}}\!\sim\!3.5$), $n_{_{\rm FIL}}\!\sim\!2\times 10^4\,{\rm cm^{-3}}$  ($n_{_4}\!\sim\!2$), and $\dot{\cal N}_{_{\rm LyC}}\!\sim\!10^{49}\,{\rm s^{-1}}$ ($\dot{\cal N}_{_{49}}\!\sim\!1$), then $\bar{R}_{_{\rm HIIR}}\!\sim\!1.5\,{\rm pc}$ and $\bar{n}_{_{\rm HIIR}}\!\sim\!400\,{\rm cm^{-3}}$; in this case, the mass-density in the H{\sc ii} region is 100 times lower than in the filament, and the H{\sc ii} region extends to roughly twice the width of the filament.

In nature, the accretion flow will not be as smooth as we have assumed, so the evolution of the H{\sc ii} region will be episodic. Moreover, as the weight of the layer of accreted gas on the boundary of the H{\sc ii} region builds up, it is likely to become Rayleigh-Taylor unstable and break up, producing a chaotic mixture of ionised gas interspersed with fingers of denser neutral gas. 

We conclude that in most cases, where a filament has spawned massive ionising stars, the ionised gas streams away rather freely, in the manner we have analysed in the preceding sections (\S\ref{SEC:Intro} through \S\ref{SEC:Approximations}). However there will be extreme cases, where the filament is relatively wide and/or relatively dense, and/or the ionising source is relatively weak. In these cases, continuing inflow onto the filament will suppress the escape of ionised gas, expansion of the H{\sc ii} region will stall, and a relatively dense, compact and turbulent H{\sc ii} region will form around the ionising stars.

\section{Discussion and Conclusions}\label{SEC:Conclusions}

We have presented a simple model for the interaction between an O star and a filament, where the O star has formed from the filament and remains close to the line defining the spine of the filament. The O star quickly ionises and disperses a section of the filament in its immediate vicinity, and the length of the dispersed section increases with time at a rate determined by the advance of the ionisation front (IF) into the exposed end of the filament. The rate of advance of the IF is determined by the fact that the amount of ionising radiation arriving at the IF is determined by how quickly the already ionised gas between the O star and IF gets out of the way. This is because most of the ionising radiation is expended maintaining ionisation against recombination in this gas. The gas immediately behind the IF is compressed by a shock front (SF), which precedes the IF into the neutral gas.

We are able to formulate simple analytic expressions for the time evolution of many of the key quantities by adopting the following rationale. There is an ionisation front with area $A_{_{\rm IF}}$, which is comparable to the cross-sectional area of the filament, $A_{_{\rm IF}}\!\sim\!\pi R_{_{\rm FIL}}^2$, and the density and speed of the gas flowing off this ionisation front are everywhere the same as on the spine of the filament (the $z$ axis). The divergence of the flow of gas off the ionisation front is characterised by the radius of curvature of the ionisation front, $R_{_{\rm IF}}$, which is comparable with the radius of the filament, $R_{_{\rm IF}}\!\sim\!R_{_{\rm FIL}}$.

With these assumptions, and adopting the dimensionless configuration parameters $R_{_{0.2}}\!\!=\!\!R_{_{\rm FIL}}/0.2{\rm pc}$, $n_{_4}\!\!=\!n_{_{\rm FIL}}/10^4{\rm cm^{-3}}$ and $\dot{\cal N}_{_{49}}\!\!=\!\!\dot{\cal N}_{_{\rm LyC}}/\!10^{49}{\rm s^{-1}}\!$, the distance from the O star to the ionisation front (measured along the axis of symmetry), $L$, the time, $t$, and the rate of advance of the IF, $dL/dt$, are given by
\begin{eqnarray}
L\!\!&\!\!\sim\!\!&\!\!5.2\,{\rm pc}\;R_{_{0.2}}^{-1/6}\,n_{_4}^{-1/3}\,\dot{\cal N}_{_{49}}^{1/6}\left[t/{\rm Myr}\right]^{2/3}\!;\\
t\!\!&\!\!\sim\!\!&\!\!0.085\,{\rm Myr}\;R_{_{0.2}}^{1/4}\,n_{_4}^{1/2}\,\dot{\cal N}_{_{49}}^{-1/4}\left[L/{\rm pc}\right]^{3/2}\!;\\
dL/dt\!\!&\!\!\sim\!\!&\!\!\left\{\!\!\begin{array}{l}
3.3\,{\rm km\,s^{-1}}\;R_{_{0.2}}^{-1/6}\,n_{_4}^{-1/3}\,\dot{\cal N}_{_{49}}^{1/6}\left[t/{\rm Myr}\right]^{-1/3}\!,\hspace{0.6cm}\\
7.5\,{\rm km\,s^{-1}}\;R_{_{0.2}}^{-1/4}\,n_{_4}^{-1/2}\,\dot{\cal N}_{_{49}}^{1/4}\left[L/{\rm pc}\right]^{-1/2}\!;\\
\end{array}\right.
\end{eqnarray}
The density of electrons at the ionisation front, $n_{_{\rm IF}}(t)$, the rate at which mass is boiled off the IF, $dM_{_{\rm II}}/dt$, the net mass of ionised gas boiled off the IF, $M_{_{\rm II}}(t)$, and the fraction of ionising photons expended maintaining the flow of ionised gas off the end of the filament, $f_{_{\rm LyC}}$, are given by
\begin{eqnarray}
n_{_{\rm IF}}\!\!&\!\!\sim\!\!&\!\!\left\{\!\!\begin{array}{l}
240\,{\rm cm^{-3}}\;R_{_{0.2}}^{-1/3}\,n_{_4}^{1/3}\,\dot{\cal N}_{_{49}}^{1/3}\left[t/{\rm Myr}\right]^{-2/3}\!,\\
1200\,{\rm cm^{-3}}\;R_{_{0.2}}^{-1/2}\,\dot{\cal N}_{_{49}}^{1/2}\left[L/{\rm pc}\right]^{-1}\!;\\
\end{array}\right.\\
dM_{_{\rm II}}/dt\!\!&\!\!\sim\!\!&\!\!\left\{\!\!\begin{array}{l}
16\,{\rm M_{_\odot}\,Myr^{-1}}\;R_{_{0.2}}^{5/3}\,n_{_4}^{1/3}\,\dot{\cal N}_{_{49}}^{1/3}\left[t/{\rm Myr}\right]^{-2/3}\!,\hspace{0.4cm}\\
83\,{\rm M_{_\odot}\,Myr^{-1}}\;R_{_{0.2}}^{3/2}\,\dot{\cal N}_{_{49}}^{1/2}\left[L/{\rm pc}\right]^{-1}\!;\\
\end{array}\right.\\
M_{_{\rm II}}\!\!&\!\!\sim\!\!&\!\!\left\{\!\!\begin{array}{l}
48\,{\rm M_{_\odot}}\;R_{_{0.2}}^{5/3}\,n_{_4}^{1/3}\,\dot{\cal N}_{_{49}}^{1/3}\left[t/{\rm Myr}\right]^{1/3}\!,\\
21\,{\rm M_{_\odot}}\;R_{_{0.2}}^{7/4}\,n_{_4}^{1/2}\,\dot{\cal N}_{_{49}}^{1/4}\left[L/{\rm pc}\right]^{1/2}\!;\\
\end{array}\right.\\
f_{_{\rm LyC}}\!\!&\!\!\sim\!\!&\!\!\left\{\!\!\begin{array}{l}
0.030\;R_{_{0.2}}^2\,\left[L/{\rm pc}\right]^{-2}\!,\\
0.11\;R_{_{0.2}}^{7/3}\;n_{_4}^{2/3}\;\dot{\cal N}_{_{49}}^{-1/3}\,\left[t/{\rm Myr}\right]^{-4/3}\!.\\
\end{array}\right.
\end{eqnarray}
The thickness of the shock-compressed layer, $W(t)$, its mass, $M_{_{\rm SCL}}(t)$, its volume-density, $n_{_{\rm SCL}}(t)$, its surface-density, $\Sigma_{_{\rm SCL}}(t)$, and its bulk velocity, $v_{_{\rm SCL}}(t)$, are given by
\begin{eqnarray}
W\!\!&\!\!\sim\!\!&\!\!\left\{\!\!\begin{array}{l}
0.18\,{\rm pc}\;R_{_{0.2}}^{13/6}\,n_{_4}^{4/3}\,\dot{\cal N}_{_{49}}^{-1/6}\left[t/{\rm Myr}\right]^{4/3}\!,\\
0.0068\,{\rm pc}\;R_{_{0.2}}^{5/2}\,n_{_4}^2\,\dot{\cal N}_{_{49}}^{-1/2}\left[L/{\rm pc}\right]^2\!;\\
\end{array}\right.\\
M_{_{\rm SCL}}\!\!&\!\!\sim\!\!&\!\!\left\{\!\!\begin{array}{l}
450\,{\rm M_{_\odot}}\;R_{_{0.2}}^{11/6}\,n_{_4}^{2/3}\,\dot{\cal N}_{_{49}}^{1/6}\left[t/{\rm Myr}\right]^{2/3}\!,\\
88\,{\rm M_{_\odot}}\;R_{_{0.2}}^2\,n_{_4}\left[L/{\rm pc}\right]\!;
\end{array}\right.\\
n_{_{\rm SCL}}\!\!&\!\!\sim\!\!&\!\!\left\{\!\!\begin{array}{l}
2.8\times 10^5\,{\rm cm^{-3}}\,R_{_{0.2}}^{-7/3}\,n_{_4}^{-5/3}\,\dot{\cal N}_{_{49}}^{1/3}\left[t/{\rm Myr}\right]^{-2/3}\!\!,\hspace{0.3cm}\\
1.4\times 10^6\,{\rm cm^{-3}}\,R_{_{0.2}}^{-5/2}\,n_{_4}^{-2}\,\dot{\cal N}_{_{49}}^{1/2}\left[L/{\rm pc}\right]^{-1}\!;\\
\end{array}\right.\\
\Sigma_{_{\rm SCL}}\!\!&\!\!\sim\!\!&\!\!\left\{\!\!\begin{array}{l}
3700\,{\rm M_{_\odot}\,pc^{-2}}\;R_{_{0.2}}^{-1/6}\,n_{_4}^{2/3}\,\dot{\cal N}_{_{49}}^{1/6}\left[t/{\rm Myr}\right]^{2/3}\!,\\
700\,{\rm M_{_\odot}\,pc^{-2}}\;n_{_4}\left[L/{\rm pc}\right]\!;\\
\end{array}\right.\\
v_{_{\rm SCL}}\!\!&\!\!\sim\!\!&\!\!\left\{\!\!\begin{array}{l}
1.9\,{\rm km\,s^{-1}}\;R_{_{0.2}}^{-1/6}\,n_{_4}^{-1/3}\,\dot{\cal N}_{_{49}}^{1/6}\left[t/{\rm Myr}\right]^{-1/3}\!,\\
5.3\,{\rm km\,s^{-1}}\;R_{_{0.2}}^{-1/4}\,n_{_4}^{-1/2}\,\dot{\cal N}_{_{49}}^{1/4}\left[L/{\rm pc}\right]^{-1/2}\!.\\
\end{array}\right.
\end{eqnarray}

These expressions are invalid if there is a very strong continuing accretion flow onto the filament. We speculate that under that circumstance, the gas ionised by the O star is trapped by the ram-pressure of the accretion flow, and the layer of neutral gas that builds up on the boundary of the H{\sc ii} region breaks up due to Rayleigh-Taylor instability, and mixes with the ionised gas.\\\\

\section*{Acknowledgements}
We thank the referee for reading our paper carefully, and pointing out where it could benefit from further discussion. APW and FDP gratefully acknowledge the support of an STFC Consolidated Grant (ST/K00926/1).

\section*{Data Availability}
There are no data in this article. All software used will be shared on request to APW.

\bibliographystyle{mnras}
\bibliography{WhitworthReferences}

\begin{thebibliography}{}
\makeatletter
\relax
\def\mn@urlcharsother{\let\do\@makeother \do\$\do\&\do\#\do\^\do\_\do\%\do\~}
\def\mn@doi{\begingroup\mn@urlcharsother \@ifnextchar [ {\mn@doi@}
  {\mn@doi@[]}}
\def\mn@doi@[#1]#2{\def\@tempa{#1}\ifx\@tempa\@empty \href
  {http://dx.doi.org/#2} {doi:#2}\else \href {http://dx.doi.org/#2} {#1}\fi
  \endgroup}
\def\mn@eprint#1#2{\mn@eprint@#1:#2::\@nil}
\def\mn@eprint@arXiv#1{\href {http://arxiv.org/abs/#1} {{\tt arXiv:#1}}}
\def\mn@eprint@dblp#1{\href {http://dblp.uni-trier.de/rec/bibtex/#1.xml}
  {dblp:#1}}
\def\mn@eprint@#1:#2:#3:#4\@nil{\def\@tempa {#1}\def\@tempb {#2}\def\@tempc
  {#3}\ifx \@tempc \@empty \let \@tempc \@tempb \let \@tempb \@tempa \fi \ifx
  \@tempb \@empty \def\@tempb {arXiv}\fi \@ifundefined
  {mn@eprint@\@tempb}{\@tempb:\@tempc}{\expandafter \expandafter \csname
  mn@eprint@\@tempb\endcsname \expandafter{\@tempc}}}

\bibitem[\protect\citeauthoryear{{Andr{\'e}} et~al.,}{{Andr{\'e}}
  et~al.}{2010}]{AndrePhetal2010}
{Andr{\'e}} P.,  et~al., 2010, \mn@doi [\aap] {10.1051/0004-6361/201014666},
  \href {https://ui.adsabs.harvard.edu/abs/2010A&A...518L.102A} {518, L102}

\bibitem[\protect\citeauthoryear{{Andr{\'e}}, {Di Francesco}, {Ward-Thompson},
  {Inutsuka}, {Pudritz}  \& {Pineda}}{{Andr{\'e}}
  et~al.}{2014}]{AndrePetal2014}
{Andr{\'e}} P.,  {Di Francesco} J.,  {Ward-Thompson} D.,  {Inutsuka} S.~I.,
  {Pudritz} R.~E.,   {Pineda} J.~E.,  2014, in {Beuther} H.,  {Klessen} R.~S.,
  {Dullemond} C.~P.,   {Henning} T.,  eds, Protostars and Planets VI. p.~27
  (\mn@eprint {arXiv} {1312.6232}),
  \mn@doi{10.2458/azu_uapress_9780816531240-ch002}

\bibitem[\protect\citeauthoryear{{Arzoumanian} et~al.,}{{Arzoumanian}
  et~al.}{2019}]{Arzoumanetal2019}
{Arzoumanian} D.,  et~al., 2019, \mn@doi [\aap] {10.1051/0004-6361/201832725},
  \href {https://ui.adsabs.harvard.edu/abs/2019A&A...621A..42A} {621, A42}

\bibitem[\protect\citeauthoryear{{Arzoumanian} et~al.,}{{Arzoumanian}
  et~al.}{2020}]{Arzoumanetal2020}
{Arzoumanian} D.,  et~al., 2020, arXiv e-prints, \href
  {https://ui.adsabs.harvard.edu/abs/2020arXiv201213060A} {p. arXiv:2012.13060}

\bibitem[\protect\citeauthoryear{{Bertoldi}}{{Bertoldi}}{1989}]{BertoldiF1989}
{Bertoldi} F.,  1989, \mn@doi [\apj] {10.1086/168055}, \href
  {https://ui.adsabs.harvard.edu/abs/1989ApJ...346..735B} {346, 735}

\bibitem[\protect\citeauthoryear{{Bisbas}, {W{\"u}nsch}, {Whitworth}, {Hubber}
  \& {Walch}}{{Bisbas} et~al.}{2011}]{BisbasTetal2011}
{Bisbas} T.~G.,  {W{\"u}nsch} R.,  {Whitworth} A.~P.,  {Hubber} D.~A.,
  {Walch} S.,  2011, \mn@doi [\apj] {10.1088/0004-637X/736/2/142}, \href
  {https://ui.adsabs.harvard.edu/abs/2011ApJ...736..142B} {736, 142}

\bibitem[\protect\citeauthoryear{{Chevance} et~al.,}{{Chevance}
  et~al.}{2020}]{ChevanceMetal2020}
{Chevance} M.,  et~al., 2020, arXiv e-prints, \href
  {https://ui.adsabs.harvard.edu/abs/2020arXiv201013788C} {p. arXiv:2010.13788}

\bibitem[\protect\citeauthoryear{{Dale}}{{Dale}}{2015}]{DaleJE2015}
{Dale} J.~E.,  2015, \mn@doi [\nar] {10.1016/j.newar.2015.06.001}, \href
  {https://ui.adsabs.harvard.edu/abs/2015NewAR..68....1D} {68, 1}

\bibitem[\protect\citeauthoryear{{Dale} \& {Bonnell}}{{Dale} \&
  {Bonnell}}{2011}]{DaleBonnell2011}
{Dale} J.~E.,  {Bonnell} I.,  2011, \mn@doi [\mnras]
  {10.1111/j.1365-2966.2011.18392.x}, \href
  {https://ui.adsabs.harvard.edu/abs/2011MNRAS.414..321D} {414, 321}

\bibitem[\protect\citeauthoryear{{Dale}, {Bonnell}  \& {Whitworth}}{{Dale}
  et~al.}{2007}]{DaleJetal2007}
{Dale} J.~E.,  {Bonnell} I.~A.,   {Whitworth} A.~P.,  2007, \mn@doi [\mnras]
  {10.1111/j.1365-2966.2006.11368.x}, \href
  {https://ui.adsabs.harvard.edu/abs/2007MNRAS.375.1291D} {375, 1291}

\bibitem[\protect\citeauthoryear{{Dale}, {Ercolano}  \& {Bonnell}}{{Dale}
  et~al.}{2012}]{DaleJEetal2012}
{Dale} J.~E.,  {Ercolano} B.,   {Bonnell} I.~A.,  2012, \mn@doi [\mnras]
  {10.1111/j.1365-2966.2012.21205.x}, \href
  {https://ui.adsabs.harvard.edu/abs/2012MNRAS.424..377D} {424, 377}

\bibitem[\protect\citeauthoryear{{Elmegreen} \& {Elmegreen}}{{Elmegreen} \&
  {Elmegreen}}{1978}]{ElmegreeBGDM1978}
{Elmegreen} B.~G.,  {Elmegreen} D.~M.,  1978, \mn@doi [\apj] {10.1086/155991},
  \href {https://ui.adsabs.harvard.edu/abs/1978ApJ...220.1051E} {220, 1051}

\bibitem[\protect\citeauthoryear{{Elmegreen} \& {Lada}}{{Elmegreen} \&
  {Lada}}{1977}]{ElmegreenLada1977}
{Elmegreen} B.~G.,  {Lada} C.~J.,  1977, \mn@doi [\apj] {10.1086/155302}, \href
  {https://ui.adsabs.harvard.edu/abs/1977ApJ...214..725E} {214, 725}

\bibitem[\protect\citeauthoryear{{Geen}, {Hennebelle}, {Tremblin}  \&
  {Rosdahl}}{{Geen} et~al.}{2015}]{GeenSetal2015}
{Geen} S.,  {Hennebelle} P.,  {Tremblin} P.,   {Rosdahl} J.,  2015, \mn@doi
  [\mnras] {10.1093/mnras/stv2272}, \href
  {https://ui.adsabs.harvard.edu/abs/2015MNRAS.454.4484G} {454, 4484}

\bibitem[\protect\citeauthoryear{{Gritschneder}, {Burkert}, {Naab}  \&
  {Walch}}{{Gritschneder} et~al.}{2010}]{GritschnederMetal2010}
{Gritschneder} M.,  {Burkert} A.,  {Naab} T.,   {Walch} S.,  2010, \mn@doi
  [\apj] {10.1088/0004-637X/723/2/971}, \href
  {https://ui.adsabs.harvard.edu/abs/2010ApJ...723..971G} {723, 971}

\bibitem[\protect\citeauthoryear{{Hacar}, {Tafalla}, {Forbrich}, {Alves},
  {Meingast}, {Grossschedl}  \& {Teixeira}}{{Hacar}
  et~al.}{2018}]{HacarAetal2018}
{Hacar} A.,  {Tafalla} M.,  {Forbrich} J.,  {Alves} J.,  {Meingast} S.,
  {Grossschedl} J.,   {Teixeira} P.~S.,  2018, \mn@doi [\aap]
  {10.1051/0004-6361/201731894}, \href
  {https://ui.adsabs.harvard.edu/abs/2018A&A...610A..77H} {610, A77}

\bibitem[\protect\citeauthoryear{{Haid}, {Walch}, {Seifried}, {W{\"u}nsch},
  {Dinnbier}  \& {Naab}}{{Haid} et~al.}{2019}]{HaidSetal2019}
{Haid} S.,  {Walch} S.,  {Seifried} D.,  {W{\"u}nsch} R.,  {Dinnbier} F.,
  {Naab} T.,  2019, \mn@doi [\mnras] {10.1093/mnras/sty2938}, \href
  {https://ui.adsabs.harvard.edu/abs/2019MNRAS.482.4062H} {482, 4062}

\bibitem[\protect\citeauthoryear{{Heitsch}, {Burkert}, {Hartmann}, {Slyz}  \&
  {Devriendt}}{{Heitsch} et~al.}{2005}]{HeitschFetal2005}
{Heitsch} F.,  {Burkert} A.,  {Hartmann} L.~W.,  {Slyz} A.~D.,   {Devriendt} J.
  E.~G.,  2005, \mn@doi [\apjl] {10.1086/498413}, \href
  {https://ui.adsabs.harvard.edu/abs/2005ApJ...633L.113H} {633, L113}

\bibitem[\protect\citeauthoryear{{Howard}, {Whitworth}, {Marsh}, {Clarke},
  {Griffin}, {Smith}  \& {Lomax}}{{Howard} et~al.}{2019}]{HowardADetal2019}
{Howard} A.~D.~P.,  {Whitworth} A.~P.,  {Marsh} K.~A.,  {Clarke} S.~D.,
  {Griffin} M.~J.,  {Smith} M.~W.~L.,   {Lomax} O.~D.,  2019, \mn@doi [\mnras]
  {10.1093/mnras/stz2234}, \href
  {https://ui.adsabs.harvard.edu/abs/2019MNRAS.489..962H} {489, 962}

\bibitem[\protect\citeauthoryear{{Kahn}}{{Kahn}}{1954}]{KahnFD1954}
{Kahn} F.~D.,  1954, \bain, \href
  {https://ui.adsabs.harvard.edu/abs/1954BAN....12..187K} {12, 187}

\bibitem[\protect\citeauthoryear{{K{\"o}nyves }, {Andr{\'e}}, {Schneider},
  {Palmeirim}, {Arzoumanian}  \& {Men'shchikov}}{{K{\"o}nyves }
  et~al.}{2013}]{KonyvesVetal2013}
{K{\"o}nyves } V.,  {Andr{\'e}} P.,  {Schneider} N.,  {Palmeirim} P.,
  {Arzoumanian} D.,   {Men'shchikov} A.,  2013, \mn@doi [Astronomische
  Nachrichten] {10.1002/asna.201211956}, \href
  {https://ui.adsabs.harvard.edu/abs/2013AN....334..908K} {334, 908}

\bibitem[\protect\citeauthoryear{{K{\"o}nyves} et~al.,}{{K{\"o}nyves}
  et~al.}{2015}]{KonyvesV2015}
{K{\"o}nyves} V.,  et~al., 2015, \mn@doi [\aap] {10.1051/0004-6361/201525861},
  \href {https://ui.adsabs.harvard.edu/abs/2015A&A...584A..91K} {584, A91}

\bibitem[\protect\citeauthoryear{{Lada}, {Lombardi}  \& {Alves}}{{Lada}
  et~al.}{2010}]{LadaCetal2010}
{Lada} C.~J.,  {Lombardi} M.,   {Alves} J.~F.,  2010, \mn@doi [\apj]
  {10.1088/0004-637X/724/1/687}, \href
  {https://ui.adsabs.harvard.edu/abs/2010ApJ...724..687L} {724, 687}

\bibitem[\protect\citeauthoryear{{Ladjelate} et~al.,}{{Ladjelate}
  et~al.}{2020}]{Ladjelatetal2020}
{Ladjelate} B.,  et~al., 2020, \mn@doi [\aap] {10.1051/0004-6361/201936442},
  \href {https://ui.adsabs.harvard.edu/abs/2020A&A...638A..74L} {638, A74}

\bibitem[\protect\citeauthoryear{{Lefloch} \& {Lazareff}}{{Lefloch} \&
  {Lazareff}}{1994}]{LeflochLazareff1994}
{Lefloch} B.,  {Lazareff} B.,  1994, \aap, \href
  {https://ui.adsabs.harvard.edu/abs/1994A&A...289..559L} {289, 559}

\bibitem[\protect\citeauthoryear{{Matzner}}{{Matzner}}{2002}]{MatznerCD2002}
{Matzner} C.~D.,  2002, \mn@doi [\apj] {10.1086/338030}, \href
  {https://ui.adsabs.harvard.edu/abs/2002ApJ...566..302M} {566, 302}

\bibitem[\protect\citeauthoryear{{Minier} et~al.,}{{Minier}
  et~al.}{2013}]{MinierVetal2013}
{Minier} V.,  et~al., 2013, \mn@doi [\aap] {10.1051/0004-6361/201219423}, \href
  {https://ui.adsabs.harvard.edu/abs/2013A&A...550A..50M} {550, A50}

\bibitem[\protect\citeauthoryear{{Peretto} et~al.,}{{Peretto}
  et~al.}{2013}]{PerettoNetal2013}
{Peretto} N.,  et~al., 2013, \mn@doi [\aap] {10.1051/0004-6361/201321318},
  \href {https://ui.adsabs.harvard.edu/abs/2013A&A...555A.112P} {555, A112}

\bibitem[\protect\citeauthoryear{{Schneider} et~al.,}{{Schneider}
  et~al.}{2012}]{Schneideretal2012}
{Schneider} N.,  et~al., 2012, \mn@doi [\aap] {10.1051/0004-6361/201118566},
  \href {https://ui.adsabs.harvard.edu/abs/2012A&A...540L..11S} {540, L11}

\bibitem[\protect\citeauthoryear{{Soler} et~al.,}{{Soler}
  et~al.}{2017}]{SolerJetal2017}
{Soler} J.~D.,  et~al., 2017, \mn@doi [\aap] {10.1051/0004-6361/201730608},
  \href {https://ui.adsabs.harvard.edu/abs/2017A&A...603A..64S} {603, A64}

\bibitem[\protect\citeauthoryear{{Tremblin}, {Audit}, {Minier}  \&
  {Schneider}}{{Tremblin} et~al.}{2012a}]{TremblinPetal2012a}
{Tremblin} P.,  {Audit} E.,  {Minier} V.,   {Schneider} N.,  2012a, \mn@doi
  [\aap] {10.1051/0004-6361/201118031}, \href
  {https://ui.adsabs.harvard.edu/abs/2012A&A...538A..31T} {538, A31}

\bibitem[\protect\citeauthoryear{{Tremblin}, {Audit}, {Minier}, {Schmidt}  \&
  {Schneider}}{{Tremblin} et~al.}{2012b}]{TremblinPetal2012b}
{Tremblin} P.,  {Audit} E.,  {Minier} V.,  {Schmidt} W.,   {Schneider} N.,
  2012b, \mn@doi [\aap] {10.1051/0004-6361/201219224}, \href
  {https://ui.adsabs.harvard.edu/abs/2012A&A...546A..33T} {546, A33}

\bibitem[\protect\citeauthoryear{{Tremblin} et~al.,}{{Tremblin}
  et~al.}{2014}]{TremblinPetal2014}
{Tremblin} P.,  et~al., 2014, \mn@doi [\aap] {10.1051/0004-6361/201423959},
  \href {https://ui.adsabs.harvard.edu/abs/2014A&A...568A...4T} {568, A4}

\bibitem[\protect\citeauthoryear{{Trevi{\~n}o-Morales}
  et~al.,}{{Trevi{\~n}o-Morales} et~al.}{2019}]{TrevinoMetal2019}
{Trevi{\~n}o-Morales} S.~P.,  et~al., 2019, \mn@doi [\aap]
  {10.1051/0004-6361/201935260}, \href
  {https://ui.adsabs.harvard.edu/abs/2019A&A...629A..81T} {629, A81}

\bibitem[\protect\citeauthoryear{{Vishniac}}{{Vishniac}}{1994}]{VishniacE1994}
{Vishniac} E.~T.,  1994, \mn@doi [\apj] {10.1086/174231}, \href
  {https://ui.adsabs.harvard.edu/abs/1994ApJ...428..186V} {428, 186}

\bibitem[\protect\citeauthoryear{{Walch}, {Whitworth}, {Bisbas}, {W{\"u}nsch}
  \& {Hubber}}{{Walch} et~al.}{2012}]{WalchSetal2012}
{Walch} S.~K.,  {Whitworth} A.~P.,  {Bisbas} T.,  {W{\"u}nsch} R.,   {Hubber}
  D.,  2012, \mn@doi [\mnras] {10.1111/j.1365-2966.2012.21767.x}, \href
  {https://ui.adsabs.harvard.edu/abs/2012MNRAS.427..625W} {427, 625}

\bibitem[\protect\citeauthoryear{{Walch}, {Whitworth}, {Bisbas}, {W{\"u}nsch}
  \& {Hubber}}{{Walch} et~al.}{2013}]{WalchSetal2013}
{Walch} S.,  {Whitworth} A.~P.,  {Bisbas} T.~G.,  {W{\"u}nsch} R.,   {Hubber}
  D.~A.,  2013, \mn@doi [\mnras] {10.1093/mnras/stt1115}, \href
  {https://ui.adsabs.harvard.edu/abs/2013MNRAS.435..917W} {435, 917}

\bibitem[\protect\citeauthoryear{{Watkins}, {Peretto}, {Marsh}  \&
  {Fuller}}{{Watkins} et~al.}{2019}]{WatkinsEetal2019}
{Watkins} E.~J.,  {Peretto} N.,  {Marsh} K.,   {Fuller} G.~A.,  2019, \mn@doi
  [\aap] {10.1051/0004-6361/201935277}, \href
  {https://ui.adsabs.harvard.edu/abs/2019A&A...628A..21W} {628, A21}

\bibitem[\protect\citeauthoryear{{Whitworth}}{{Whitworth}}{1979}]{WhitworthA1979}
{Whitworth} A.,  1979, \mn@doi [\mnras] {10.1093/mnras/186.1.59}, \href
  {https://ui.adsabs.harvard.edu/abs/1979MNRAS.186...59W} {186, 59}

\bibitem[\protect\citeauthoryear{{Whitworth}, {Bhattal}, {Chapman}, {Disney}
  \& {Turner}}{{Whitworth} et~al.}{1994a}]{Whitetal1994a}
{Whitworth} A.~P.,  {Bhattal} A.~S.,  {Chapman} S.~J.,  {Disney} M.~J.,
  {Turner} J.~A.,  1994a, \mn@doi [\mnras] {10.1093/mnras/268.1.291}, \href
  {https://ui.adsabs.harvard.edu/abs/1994MNRAS.268..291W} {268, 291}

\bibitem[\protect\citeauthoryear{{Whitworth}, {Bhattal}, {Chapman}, {Disney}
  \& {Turner}}{{Whitworth} et~al.}{1994b}]{Whitetal1994b}
{Whitworth} A.~P.,  {Bhattal} A.~S.,  {Chapman} S.~J.,  {Disney} M.~J.,
  {Turner} J.~A.,  1994b, \aap, \href
  {https://ui.adsabs.harvard.edu/abs/1994A&A...290..421W} {290, 421}

\bibitem[\protect\citeauthoryear{{Williams}, {Peretto}, {Avison},
  {Duarte-Cabral}  \& {Fuller}}{{Williams} et~al.}{2018}]{WilliamsG2018}
{Williams} G.~M.,  {Peretto} N.,  {Avison} A.,  {Duarte-Cabral} A.,   {Fuller}
  G.~A.,  2018, \mn@doi [\aap] {10.1051/0004-6361/201731587}, \href
  {https://ui.adsabs.harvard.edu/abs/2018A&A...613A..11W} {613, A11}

\makeatother
\end{thebibliography}

\appendix
\section{Gas velocities relative to a D-critical ionisation front}

On the basis of stability considerations, \citet{KahnFD1954} shows that, unless the neutral gas encountered by an IF is very rarefied, the IF is preceded by an SF which compresses the neutral gas so that the IF is D-critical; the neutral gas then flows into the IF subsonically, and the ionised gas flows out of the IF sonically. The density of molecular hydrogen flowing into the IF is $n_{_{\rm SCL}}$, and it flows in at speed $u_{_{\rm SCL}}\,(\!<\!c_{_{\rm SCL}}\!)$. The density of protons flowing out of the IF is $n_{_{\rm II}}$, and they flow out at speed $c_{_{\rm II}}$. Conservation of mass and momentum therefore require
\begin{eqnarray}\label{EQN:mass.1}
2n_{_{\rm SCL}}u_{_{\rm SCL}}&=&n_{_{\rm II}}c_{_{\rm II}},\\\label{EQN:momentum.1}
2n_{_{\rm SCL}}\left[c_{_{\rm SCL}}^2+u_{_{\rm SCL}}^2\right]&=&2n_{_{\rm II}}c_{_{\rm II}}^2.
\end{eqnarray}
Eliminating $n_{_{\rm SCL}}/n_{_{\rm II}}$ between these Eqs. \ref{EQN:mass.1} and \ref{EQN:momentum.1}, we obtain a quadratic equation for $u_{_{\rm SCL}}$,
\begin{eqnarray}
u_{_{\rm SCL}}^2\,-\,2c_{_{\rm II}}u_{_{\rm SCL}}\,+\,c_{_{\rm SCL}}^2&=&0,
\end{eqnarray}
with roots
\begin{eqnarray}\label{EQN:uSCL.1}
u_{_{\rm SCL}}&=&c_{_{\rm II}}\left\{1\,\pm\,\left[1\,-\,c_{_{\rm SCL}}^2/c_{_{\rm II}}^2\right]^{1/2}\right\}\\\label{EQN:uSCL.2}
&\approx&c_{_{\rm SCL}}^2/2c_{_{\rm II}}.
\end{eqnarray}
In obtaining the final result (Eq. \ref{EQN:uSCL.2}), we have set `$\pm$' to `$-$' in Eq. \ref{EQN:uSCL.1}, in order to obtain subsonic $u_{_{\rm SCL}}$ (i.e. $u_{_{\rm SCL}}\!<\!c_{_{\rm SCL}}$), and we have also only retained the highest-order finite term in $c_{_{\rm SCL}}/c_{_{\rm II}}$. Substituting $u_{_{\rm SCL}}=dL/dt-v_{_{\rm SCL}}$ we obtain Eq. \ref{EQN:dLdt.1}.

\bsp	
\label{lastpage}
\end{document}